\documentclass[a4paper,11pt]{article}

\usepackage{jheppub} 

\usepackage[T1]{fontenc} 

\title{Standard model contribution to the electric dipole moment of the deuteron, $^3$H, and $^3$He nuclei}

\author[a,1]{Nodoka Yamanaka\note{Corresponding author.}}
\author[b]{and Emiko Hiyama,}

\affiliation[a]{iTHES Research Group, RIKEN,\\
Wako, Saitama 351-0198, Japan}
\affiliation[b]{RIKEN Nishina Center, RIKEN,\\
Wako, Saitama 351-0198, Japan}

\emailAdd{nodoka.yamanaka@riken.jp}
\emailAdd{hiyama@riken.jp}

\abstract{
We calculate for the first time the electric dipole moment (EDM) of the deuteron, $^3$H, and $^3$He nuclei generated by the one-meson exchange CP-odd nuclear force in the standard model.
The effective $|\Delta S| = 1$ four-quark operators are matched to the $|\Delta S| = 1$ standard model processes involving the CP phase of the Cabibbo-Kobayashi-Maskawa matrix at the electroweak scale and run down to the hadronic scale $\mu = 1$ GeV according to the renormalization group evolution in the next-to-leading logarithmic order.
At the hadronic scale, the hadron matrix elements are modeled in the factorization approach.
We then obtain the one-meson (pion, eta meson, and kaon) exchange CP-odd nuclear force, which is the combination of the $|\Delta S| = 1$ meson-baryon vertices which issue from the penguin operator and the hyperon-nucleon transition.
From this CP-odd nuclear force, the nuclear EDM is calculated with the realistic Argonne $v18$ interaction and the CP-odd nuclear force using the Gaussian expansion method.
It is found that the EDMs of light nuclear systems are of order $O(10^{-31}) e $ cm.
We also estimate the standard model contribution to other hadronic CP violating observables such as the EDMs of $^6$Li, $^9$Be nuclei, and the atomic EDMs of $^{129}$Xe, $^{199}$Hg, $^{211}$Rn, and $^{225}$Ra generated through the nuclear Schiff moment.
We then analyze the source of theoretical uncertainties and show some possible ways to overcome them.
}

\begin{document} 
\maketitle
\flushbottom

\section{Introduction}
\label{sec:intro}

One of the most promising experimental approaches to discover the CP violation beyond the standard model (SM) is the measurement of the {\it electric dipole moment} (EDM) \cite{hereview,Bernreuther,khriplovichbook,ginges,pospelovreview,fukuyama,naviliat,engel,yamanakabook,hewett,roberts}.
The CP violation beyond SM has long been required to explain the matter excess over the antimatter of the Universe \cite{sakharov,shaposhnikov1,shaposhnikov2}.

To unveil the CP violation of the new physics through the EDM, the SM contribution must be known.
The SM effect, generated by the CP phase of the Cabibbo-Kobayashi-Maskawa (CKM) matrix \cite{ckm}, is generally very small for the EDM, and this fact makes one of the attractive feature of the EDM as an experimental probe.
The EDM of elementary particles in the SM is suppressed due to the GIM mechanism \cite{gim}.
The quark EDM appears only from the 3-loop level \cite{shabalinquarkedm,shabalinquarkedm2,smtheta,czarnecki} with the estimated value for the light quark $d_q \sim 10^{-35}e $ cm, and the charged lepton EDM from the 4-loop level \cite{smelectronedm,smelectronedm2} with $d_e \sim 10^{-41} e $ cm for the electron.
These values are well below the current experimental sensitivity \cite{rosenberry,regan,baker,muong-2,griffith,hudson,acme,parker}.

For hadronic systems, the situation is more involved.
We have to note that the hadronic EDM receives contribution from the single quark EDM and also from the CP violating many-body interactions.
For the neutron EDM, the effect of the quark EDM are found to be subleading, and the long range pion loop contribution \cite{smneutronedm,smneutronedmmckellar,seng} as well as the higher dimensional many-quark operator process \cite{mannel} are thought to be dominant.

As a more complicated system, we have the nucleus.
Recently, new technology to measure the nuclear EDM using the storage ring is being developed \cite{hewett,storage1,storage2,storage3,storage4,storage5,bnl}.
The most attractive feature of the nuclear EDM measurement using storage rings is its prospective sensitivity which may reach $O(10^{-29}) e$ cm.
This approach is much more sensitive on the hadronic CP violation than the extensively studied atomic EDM \cite{rosenberry,griffith,hudson,acme,parker}, where the hadronic CP-odd effect is screened by the atomic electrons \cite{schiff,ginges}.
Moreover, it has been shown in a recent paper that the nuclear many-body effect enhances the EDM for light nuclei \cite{yamanakanuclearedm}.
The nuclear EDM is thus one of the main target of the next generation search for the CP violation of new physics beyond SM.
Many theoretical investigations of the EDM of light nuclei were done in this context \cite{yamanakanuclearedm,
avishai1,avishai2,avishai3,
korkin,liu,pospelovdeuteron,afnan,devries,dedmtheta,stetcu,chiral3nucleon,song,Dekens:2014jka,bsaisou,bsaisou2,mereghetti,devries2,devries3}.
The determination of the SM background is therefore of crucial importance.

The SM contribution  to the nuclear EDM was discussed many times in the past.
The first attempt was done in Ref. \cite{sushkov} (see also Refs. \cite{sushkov2,sushkov3}), where the {\it penguin diagram} \cite{shifman,ellis} was considered in the simple factorization approach, and it was found that the pion pole significantly enhances the nuclear EDM.
But subsequently, it was pointed in Ref. \cite{donoghue} that this enhancement cancels due to additional contributions from the chiral condensate, respecting the chiral symmetry.
The most recent analysis has investigated the CP-odd nuclear force in the chiral approach, by introducing the CP violation of the CKM matrix as a phase of the weak spurion of the chiral $SU(3)$ group \cite{smCPVNN}.
There the CP-odd nuclear force was given as a contact nucleon-nucleon interaction.
The simple dimensional estimation of the nuclear EDM yields $d_A \sim O(\frac{\alpha_s}{4\pi} G_F^2 J \Lambda_{\rm QCD}^3 ) \sim 10^{-32}e$ cm, where $J = (3.06^{+0.21}_{-0.20} ) \times 10^{-5}$ is the Jarlskog invariant \cite{pdg,jarlskog}.

Since then, the nuclear EDM was long thought to be small, and its value has never been seriously questioned.
In the mean time, there was a significant development on how to treat the flavor changing CP violation of the CKM matrix in the context of its search in the $K$ and $B$ meson decays, especially in the perturbative treatment of the renormalization group evolution \cite{inami-lim,burasreview,buraslecture,burasleading,buras,ewpenguin1,ewpenguin2,anatomy,fleischer,keum,anatomyb,NLOweakdecay,burasnewphysics,two-loop_penguin,buras2015}.
There it was argued that the effect of the penguin diagram is sufficiently enhanced in the next-to-leading logarithmic order at low energy scale \cite{buras}.

As another significant progress, we have the ab initio calculations of the EDM of light nuclei \cite{liu,stetcu,devries,song,bsaisou,bsaisou2,yamanakanuclearedm}.
The authors have previously shown that the few-body calculations using the Gaussian expansion method \cite{hiyama,Hiyama2012ptep} work well in the ab initio evaluation of the nuclear EDM \cite{yamanakanuclearedm}.
The Gaussian expansion method is a framework which can solve the Schr\"{o}dinger equation of few-body systems, and has been successful in describing many systems of many fields \cite{kamimuramuonic,
3nucleon,
hiyamahypernuclei1,
hiyamahypernuclei2,
benchmark,
hiyamahyperon-nucleon,
hiyama1stexcitedstate,
hiyamapentaquark,
hamaguchi,
funaki,
hiyamahypernuclei3,
hiyamaatom1,
hiyamaatom2,
yokota,
kusakabe,
yamada,maeda}.
Due to the flavor change in the CP violation of the CKM matrix, the nuclear EDM in the SM is generated by the CP-odd nuclear force with strange hadrons in the intermediate state.
The nuclear EDM with kaon exchange CP-odd nuclear force has never been evaluated previously.
This problem can be resolved using the Gaussian expansion method.

From the above arguments, the evaluation of the nuclear EDM in the SM seems quite timely.
In this work, we therefore predict the nuclear EDM of the deuteron, $^3$H, and $^3$He within the SM. 
We also estimate the EDM of other nuclear systems.
Let us recapitulate the important improvements which will be evaluated in our calculation.
First, the renormalization group evolution of the $|\Delta S|=1$ four-quark operators will be calculated.
This short range strong interaction effect is very important to consider because the Wilson coefficients of the penguin operators are enhanced by more than an order of magnitude when we move from the electroweak scale to the hadronic scale.
Next, we establish the formulae of the P, CP-odd nuclear couplings when the strangeness is shifted in the intermediate states.
The derivation of the CP-odd nuclear force with several pseudoscalar mesons in the factorization approach is given for the first time.
Within the factorization model, our calculation suggests an enhancement for the kaon exchange CP-odd nuclear coupling, due to the large value of the isoscalar nucleon scalar density (pion-nucleon sigma term).
Finally, the importance of this $K$ meson exchange CP violation motivates us to evaluate for the first time the EDM of light nuclei in the presence of this CP-odd nuclear force in the Gaussian expansion method.
This contribution has never been evaluated beyond the contact interaction approximation.
As the pion exchange effect is large at the nuclear level, those two meson exchange processes may be of the same order of magnitude, and the accurate evaluation of the relative size is therefore crucial.
It is also important to note that those procedures can also be applied to the analysis of models beyond SM with flavor violation, although we do not do it in this work.

This paper is organized as follows.
We first match the effective $|\Delta S|=1$ four-quark interactions to the elementary level perturbative processes at the electroweak scale, and run them down to the hadronic scale using the renormalization group equation.
We then calculate the hadron level effective $|\Delta S|=1$ interaction in the factorization approach, and derive the pion, eta meson, and kaon exchange CP-odd nuclear forces within chiral symmetry-inspired model.
From the obtained CP-odd nuclear force, we calculate the EDM of the deuteron, $^3$H, and $^3$He using the Gaussian expansion method.
For that, we consider the $K$ meson exchange CP-odd nuclear force for the first time.
We also estimate the EDM of $^6$Li, $^9$Be nuclei, and the nuclear Schiff moment contribution to the EDM of several atoms within the pion exchange CP-odd nuclear force.
After presenting the result of our calculation, we analyze the sources of theoretical uncertainties, and the way to overcome them in the future. 
We finally summarize the discussion.

\section{Calculation of the nuclear EDM}

\subsection{The renormalization of the $|\Delta S| =1$ four-quark interactions}

In this section, we derive the effective $|\Delta S| =1$ four-quark interactions renormalized at the hadronic scale starting from the elementary particle level physics.
The derivation follows Ref. \cite{buras} in which the renormalization of the $|\Delta S| =1$ four-quark interaction in the next-to-leading logarithmic approximation was discussed.

\subsubsection{Matching of the elementary level process at $\mu = m_W$}

\begin{figure}[htb]
\begin{center}
\includegraphics[width=12cm]{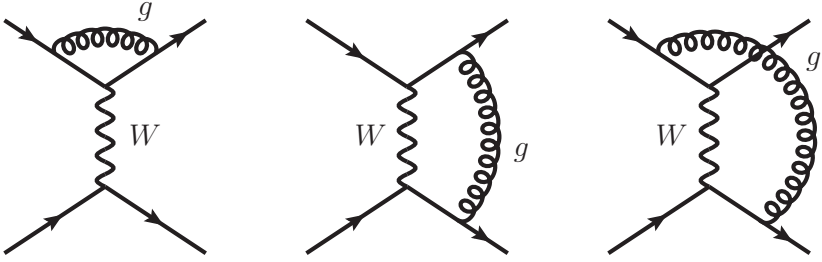}
\caption{\label{fig:current-current_correction}
QCD corrections to the $|\Delta S| = 1$ $W$ boson exchange diagram.
The solid lines indicate the quarks.
Other diagrams with the up-down or left-right reflections also contribute.
}
\end{center}
\end{figure}

\begin{figure}[htb]
\begin{center}
\includegraphics[width=5cm]{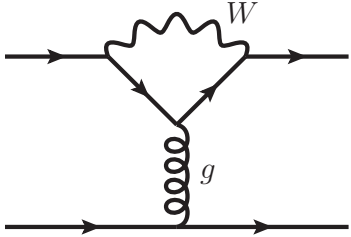}
\caption{\label{fig:penguin}
The gluonic penguin diagram.
The solid lines denote the quarks.
}
\end{center}
\end{figure}

We first match the $|\Delta S| = 1$ SM processes involving the CP phase of the CKM matrix with the effective $|\Delta S| = 1$ four-quark operators at the scale of the $W$ boson mass $\mu = m_W= 80.385 \pm 0.015$ GeV \cite{pdg}.
The SM contributions to the $|\Delta S| = 1$ four-quark processes are given in Figs. \ref{fig:current-current_correction} and \ref{fig:penguin}.
In this perturbative evaluation, the strong coupling $\alpha_s$ must be fixed at the scale $\mu = m_W= 80.385$ GeV.
We have used $\alpha_s (m_Z) = 0.1185$ as input \cite{pdg}, and found $\alpha_s (m_W) = 0.01208$
 after two-loop level renormalization \cite{tarasov}.
The top quark mass in the $\overline{\rm MS}$ scheme $m_t = 160$ GeV is fixed for theories below the scale $\mu = m_t$.
It should be noted that the top quark mass must be renormalized in the $\overline{\rm MS}$ scheme, to be consistent with the scheme of the renormalization group evolution adopted in this work \cite{smith}.

The $|\Delta S| =1$ effective hamiltonian we want to match is
\begin{eqnarray}
{\cal H}_{eff} (\mu = m_W)
&=&
\frac{G_F}{\sqrt{2}}
\Biggl\{
C_1 (\mu = m_W) [ V_{us}^* V_{ud} Q_1^u + V_{cs}^* V_{cd} Q_1^c ]
\nonumber\\
&& \hspace{3em}
+C_2 (\mu = m_W) [ V_{us}^* V_{ud} Q_2^u + V_{cs}^* V_{cd} Q_2^c ]
\nonumber\\
&& \hspace{8em}
- V_{ts}^* V_{td} \sum_{i=3}^6 C_i (\mu = m_W) Q_i
\Biggr\}
+{\rm h.c.}
,
\label{eq:effhamimw}
\end{eqnarray}
where $V_{qq'}$ are the CKM matrix elements and $G_F = 1.16637 \times 10^{-5} {\rm GeV}^{-2}$ is the Fermi constant \cite{pdg}.
The $\Delta S =-1$ four-quark operator basis is given by \cite{buras,burasreview,buraslecture}
\begin{eqnarray}
Q_1^q &=&
\bar s_\alpha \gamma^\mu (1-\gamma_5) q_\beta \cdot \bar q_\beta \gamma_\mu (1-\gamma_5) d_\alpha
,
\label{eq:q1}
\\
Q_2^q &=&
\bar s_\alpha \gamma^\mu (1-\gamma_5) q_\alpha \cdot \bar q_\beta \gamma_\mu (1-\gamma_5) d_\beta
,
\label{eq:q2}
\\
Q_3 &=&
\bar s_\alpha \gamma^\mu (1-\gamma_5) d_\alpha \cdot \sum_q^{N_f} \bar q_\beta \gamma_\mu (1-\gamma_5) q_\beta
,
\label{eq:q3}
\\
Q_4 &=&
\bar s_\alpha \gamma^\mu (1-\gamma_5) d_\beta \cdot \sum_q^{N_f} \bar q_\beta \gamma_\mu (1-\gamma_5) q_\alpha
,
\label{eq:q4}
\\
Q_5 &=&
\bar s_\alpha \gamma^\mu (1-\gamma_5) d_\alpha \cdot \sum_q^{N_f} \bar q_\beta \gamma_\mu (1+\gamma_5) q_\beta
,
\label{eq:q5}
\\
Q_6 &=&
\bar s_\alpha \gamma^\mu (1-\gamma_5) d_\beta \cdot \sum_q^{N_f} \bar q_\beta \gamma_\mu (1+\gamma_5) q_\alpha
,
\label{eq:q6}
\end{eqnarray}
where $\alpha$ and $\beta$ are the color indices of the fundamental representation.
At the scale $\mu = m_W$, the number of active flavors is $N_f =5$.
In this work, we do not consider the electroweak penguin contributions which require additional operators \cite{ewpenguin1,ewpenguin2,anatomy}.

\begin{figure}[htb]
\begin{center}
\includegraphics[width=12cm]{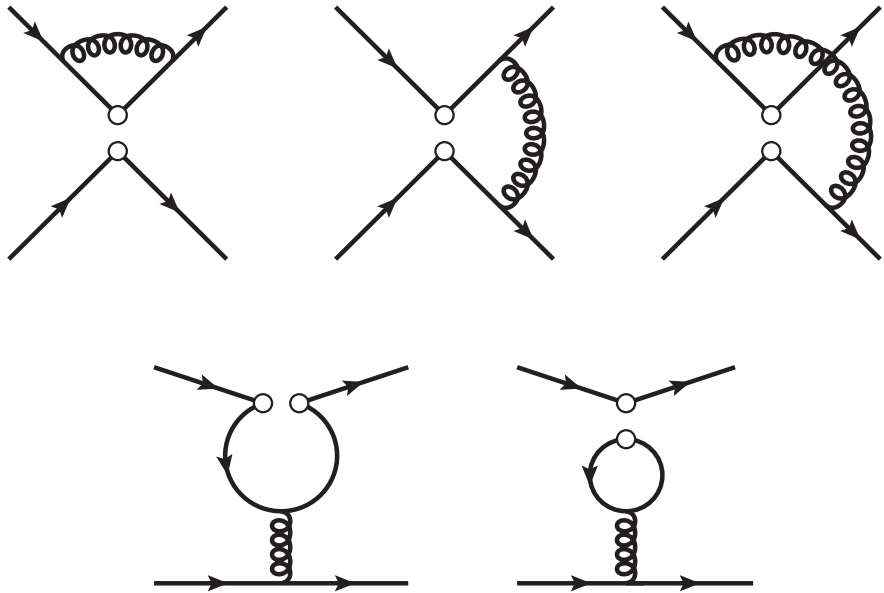}
\caption{\label{fig:effective_interaction_correction}
QCD corrections to the $|\Delta S| = 1$ effective four-quark interaction.
The solid lines and the curly lines denote the quarks and the gluons, respectively.
The white double blobs give the effective $|\Delta S| =1$ four-quark vertices.
Other diagrams with the up-down or left-right reflections also contribute.
}
\end{center}
\end{figure}

To match the amplitude of the diagrams of Figs. \ref{fig:current-current_correction} and \ref{fig:penguin} with the amplitude generated by the effective hamiltonian (\ref{eq:effhamimw}) consistently in the order of $\alpha_s$, we also have to consider the QCD corrections to the latter.
The QCD corrections are given in Fig. \ref{fig:effective_interaction_correction}.
We see that the operators $Q_3$, $Q_4$, $Q_5$, and $Q_6$ are generated by the penguin diagram (see Fig. \ref{fig:penguin}) with the top quark.

After matching, we obtain the following Wilson coefficients to the next-to-leading order $|\Delta S| =1$ four-quark interaction:
\begin{equation}
{\bf C} (\mu = m_W)
\equiv
\left(
\begin{array}{c}
C_1 (\mu = m_W) \cr
C_2 (\mu = m_W) \cr
C_3 (\mu = m_W) \cr
C_4 (\mu = m_W) \cr
C_5 (\mu = m_W) \cr
C_6 (\mu = m_W) \cr
\end{array}
\right)
=
\left(
\begin{array}{c}
\frac{11}{2} \frac{\alpha_s (m_W)}{4\pi} \cr
1-\frac{11}{6} \frac{\alpha_s (m_W)}{4\pi} \cr
-\frac{\alpha_s (m_W)}{24\pi} \tilde E (x_t) \cr
\frac{\alpha_s (m_W)}{8\pi} \tilde E (x_t) \cr
-\frac{\alpha_s (m_W)}{24\pi} \tilde E (x_t) \cr
\frac{\alpha_s (m_W)}{8\pi} \tilde E (x_t) \cr
\end{array}
\right)
=
\left(
\begin{array}{c}
5.29 \times 10^{-2} \cr
9.82 \times 10^{-1} \cr
6.19 \times 10^{-4} \cr
-1.86 \times 10^{-3} \cr
6.19 \times 10^{-4} \cr
-1.86 \times 10^{-3} \cr
\end{array}
\right)
,
\label{eq:wilsoncoefinit}
\end{equation}
where 
\begin{equation}
\tilde E (x_t ) 
= 
- \frac{2}{3}
- \frac{2}{3} \ln x_t
+ \frac{x_t^2 (15 - 16x_t +4 x_t^2) }{6 (1-x_t)^4} \ln x_t
+ \frac{x_t (18 -11 x_t -x_t^2)}{12(1-x_t)^3}
,
\end{equation}
with $x_t \equiv \frac{m_t^2}{m_W^2}$.
Note that the Wilson coefficients depend on the renormalization scheme.
Here we have adopted the naive dimensional regularization scheme.
We see that the coefficient of the operator $Q_2$ is the largest, due to the tree level contribution.
The Wilson coefficients of Eq. (\ref{eq:wilsoncoefinit}) give the initial condition for the renormalization group evolution.

\subsubsection{Renormalization of four-quark interactions down to the hadronic scale}

We now run the $|\Delta S|=1$ four-quark interactions down to the hadronic scale $\mu = 1$ GeV according to the renormalization group evolution in the next-to-leading logarithmic order \cite{buras,burasreview}.
The evolution of the Wilson coefficients in the change of the renormalization point can be written as
\begin{equation}
\frac{d}{d \ln \mu} {\mathbf C} ( \mu )
=
\hat \gamma^T (\alpha_s) {\mathbf C} ( \mu )
,
\label{eq:renor_eq}
\end{equation}
where the anomalous dimension matrix $\hat \gamma$ is defined by
\begin{equation}
\hat \gamma
=
Z^{-1}
\frac{d}{d \ln \mu}
Z
,
\end{equation}
with $Z$ the renormalization factor (matrix) relating the bare and renormalized Wilson coefficients, as ${\mathbf C}^{(0)} = Z {\mathbf C} ( \mu )$.
By integrating (\ref{eq:renor_eq}) and applying the initial condition at the scale $\mu' = m_W$, given by Eq. (\ref{eq:wilsoncoefinit}), we obtain
\begin{equation}
{\mathbf C} ( \mu )
=
\hat{U} ( \mu , \mu' = m_W)
{\mathbf C} (  \mu' = m_W)
,
\end{equation}
where
\begin{equation}
\hat{U} ( \mu , \mu' =  m_W)
=
T_g
\exp
\int_{g(m_W)}^{g(\mu)} dg' \frac{\gamma^T (g')}{\beta (g')}
,
\end{equation}
with the QCD coupling $g \equiv \sqrt{4\pi \alpha_s}$.
The anomalous dimension matrix and the beta function are expanded in terms of the strong coupling as
\begin{eqnarray}
\hat \gamma (g)
&=&
\hat \gamma^{(0)} \frac{g^2}{16\pi^2}
+ \hat \gamma^{(1)} \frac{g^4}{(16\pi^2)^2}
+\cdots
,
\\
\beta (g)
&=&
-\beta_0 \frac{g^3}{16 \pi^2}
-\beta_1 \frac{g^5}{(16 \pi^2)^2}
+\cdots
,
\end{eqnarray}
where we take only the first two terms in the next-to-leading logarithmic order.
The coefficients of the beta function $\beta_0 = 11 - \frac{2}{3} N_f$ and $\beta_1 =102 - \frac{38}{3} N_f$ are well known for $N_c =3$.
The anomalous dimension matrices $\hat \gamma^{(0)}$ and $\hat \gamma^{(1)}$, also depending on $N_f$, are given in Appendix \ref{sec:penguinrge}.

The calculation of the evolution of the effective hamiltonian (\ref{eq:effhamimw}) needs care, since the matching changes its form after crossing each quark mass threshold.
The evolution of the Wilson coefficient array down to the hadronic scale $m_b > \mu > m_c$ is given by 
\begin{equation}
{\mathbf C} (m_b > \mu > m_c)
=
\hat{U}_4 ( \mu , m_b) \hat{R} \, \hat{U}_5 ( m_b , m_W ) {\mathbf C} (\mu = m_W)
.
\end{equation}
Here $\hat{U}_{N_f}$ is the running matrix with the active number of flavor $N_f$, calculated at the next-to-leading order \cite{buras}.
When we cross the scale $\mu = m_b = 4.18$ GeV \cite{pdg}, the bottom quark becomes inactive and $N_f$ changes from 5 to 4.
There we have to match the effective theory with $N_f = 5$ to that $N_f =4$, where the penguin operators $Q_3$, $Q_4$, $Q_5$, and $Q_6$ are changed.
The matching is systematically achieved by multiplying the quark threshold matching matrix $\hat{R}$.
The explicit expressions of the matrices $\hat{U}_{N_f}$ and $\hat{R}$ are given in Appendix \ref{sec:penguinrge}.
For the scale above the charm quark mass $\mu > m_c$, the evolution of the four-quark operators $Q_1^u$ and $Q_1^c$ is described by a common Wilson coefficient (the same remark applies for $Q_2^u$ and $Q_2^c$).
Let us show the Wilson coefficients for $\mu = 2$ GeV, which is a renormalization point often used:
\begin{equation}
{\bf C} ( \mu =2\, {\rm GeV} )
=
\left(
\begin{array}{c}
-0.313 \cr
1.15 \cr
1.98 \times 10^{-2} \cr
-4.56 \times 10^{-2}  \cr
1.24 \times 10^{-2}  \cr
-6.27 \times 10^{-2} \cr
\end{array}
\right)
.
\end{equation}

At the scale lower than $\mu = m_c = 1.275$ GeV \cite{pdg}, the effective $|\Delta S| = 1$ four-quark hamiltonian looses operators involving the charm quark field and becomes \cite{vainshtein1,flynn,burasleading,paschos,bardeen,gilman,guberina}
\begin{equation}
{\cal H}_{eff} (\mu)
=
\frac{G_F}{\sqrt{2}} V_{us}^* V_{ud}
\sum_{i=1}^6
[ \tilde z_i (\mu) + \tau \tilde y_i (\mu)] Q_i (\mu)
+ {\rm h.c.}
,
\label{eq:effhamibelowmc}
\end{equation}
where $\tau \equiv - \frac{V_{ts}^* V_{td}}{V_{us}^* V_{ud}}$.
Below $\mu = m_c$, we have $Q_1 = Q_1^u$ and $Q_2 = Q_2^u$.
The Wilson coefficients $\tilde z_i$ are given by
\begin{equation}
\tilde {\bf z} ( \mu < m_c )
=
\hat{U}_3 ( \mu , m_c) 
\left(
\begin{array}{c}
C_1 (m_c) \cr
C_2 (m_c) \cr
-\frac{\alpha_s (m_c)}{24\pi} \tilde F (m_c) \cr
\frac{\alpha_s (m_c)}{8\pi} \tilde F (m_c) \cr
-\frac{\alpha_s (m_c)}{24\pi} \tilde F (m_c) \cr
\frac{\alpha_s (m_c)}{8\pi} \tilde F (m_c) \cr
\end{array}
\right)
,
\end{equation}
where $\tilde F (m_c) = -\frac{2}{3} C_2 (m_c)$ in the naive dimensional regularization \cite{anatomy}.
The array of the above equation is the result of the matching at the scale of charm quark mass.
The first two Wilson coefficients issue from the standard renormalization group evolution, since the operators $Q_1^u$ and $Q_2^u$ are inherited below the scale $\mu = m_c$.
The operators $Q_1^c$ and $Q_2^c$ are integrated out for $\mu < m_c$.
Therefore penguin operators are generated due to the lower diagrams of Fig. \ref{fig:effective_interaction_correction}, and the Wilson coefficients $\tilde z_i$ $(i=3\sim 6)$ become nonzero.
Here we have used the CKM unitarity
\begin{equation}
V_{cs}^* V_{cd} 
=
-V_{us}^* V_{ud} - V_{ts}^* V_{td} 
.
\label{eq:ckmunitarity}
\end{equation}

The Wilson coefficients $\tilde y_i$ at $\mu = m_c$ receive contribution from the Wilson coefficients evolved from the initial penguin operators at $\mu = m_W$ [last line of Eq. (\ref{eq:effhamimw})], and from the integration of the charm quark at $\mu = m_c$, due to the last term of the CKM unitarity (\ref{eq:ckmunitarity}).
Their expression at the scale $\mu < m_c$ is explicitly given by
\begin{equation}
\tilde y_i (\mu)
=
\tilde v_i (\mu) - \tilde z_i (\mu)
,
\end{equation}
where
\begin{equation}
\tilde {\mathbf v} (\mu )
=
\hat{U}_3 (\mu , m_c) \hat{R} \, \hat{U}_4 ( m_c , m_b) \hat{R} \, \hat{U}_5 ( m_b , m_W ) {\mathbf C} (\mu = m_W)
.
\end{equation}
We can remark that the Wilson coefficients $\tilde y_1$ and $\tilde y_2$ are vanishing.

\begin{figure}[htbp]
\begin{minipage}{0.5\hsize}

\begin{center}
\includegraphics[width=7.5cm]{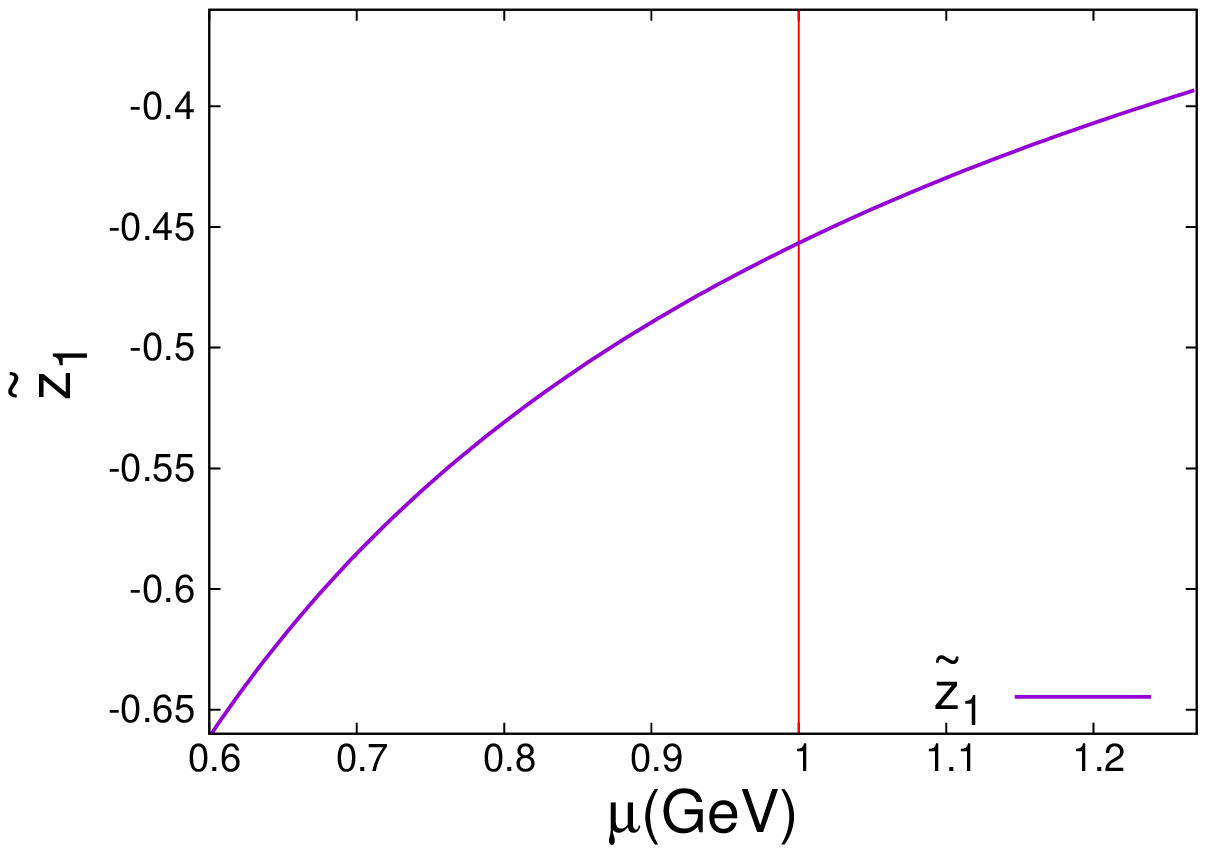}
\end{center}

\end{minipage}
\begin{minipage}{0.5\hsize}
 
\begin{center}
\includegraphics[width=7.5cm]{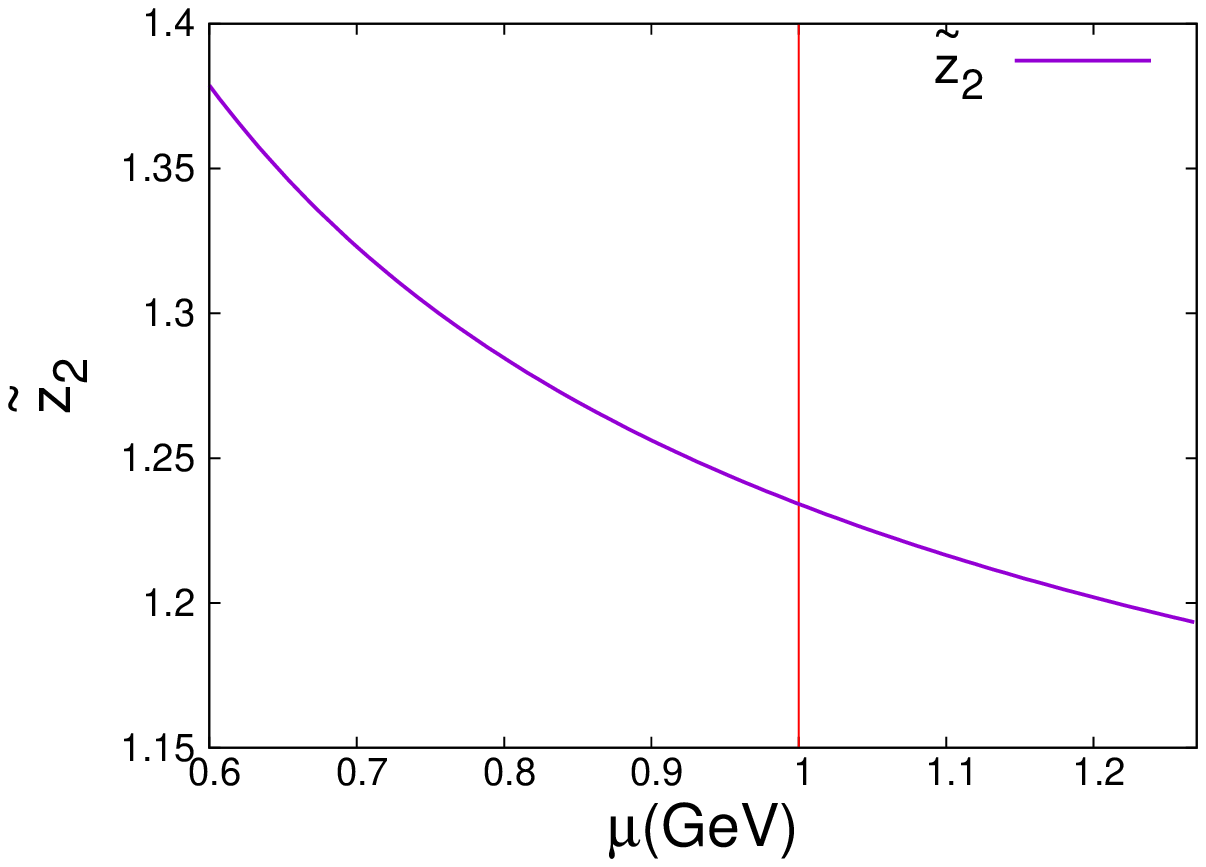}
\end{center}

\end{minipage}
\caption{\label{fig:ztilde}
The behavior of the Wilson coefficients $\tilde z_1$ and $\tilde z_2$ for the scale below the charm quark mass.
The red vertical line denotes the hadronic renormalization scale ($\mu =1$ GeV) we have chosen.
}
\end{figure}

\begin{figure}[htbp]
\begin{minipage}{0.5\hsize}

\begin{center}
\includegraphics[width=7.5cm]{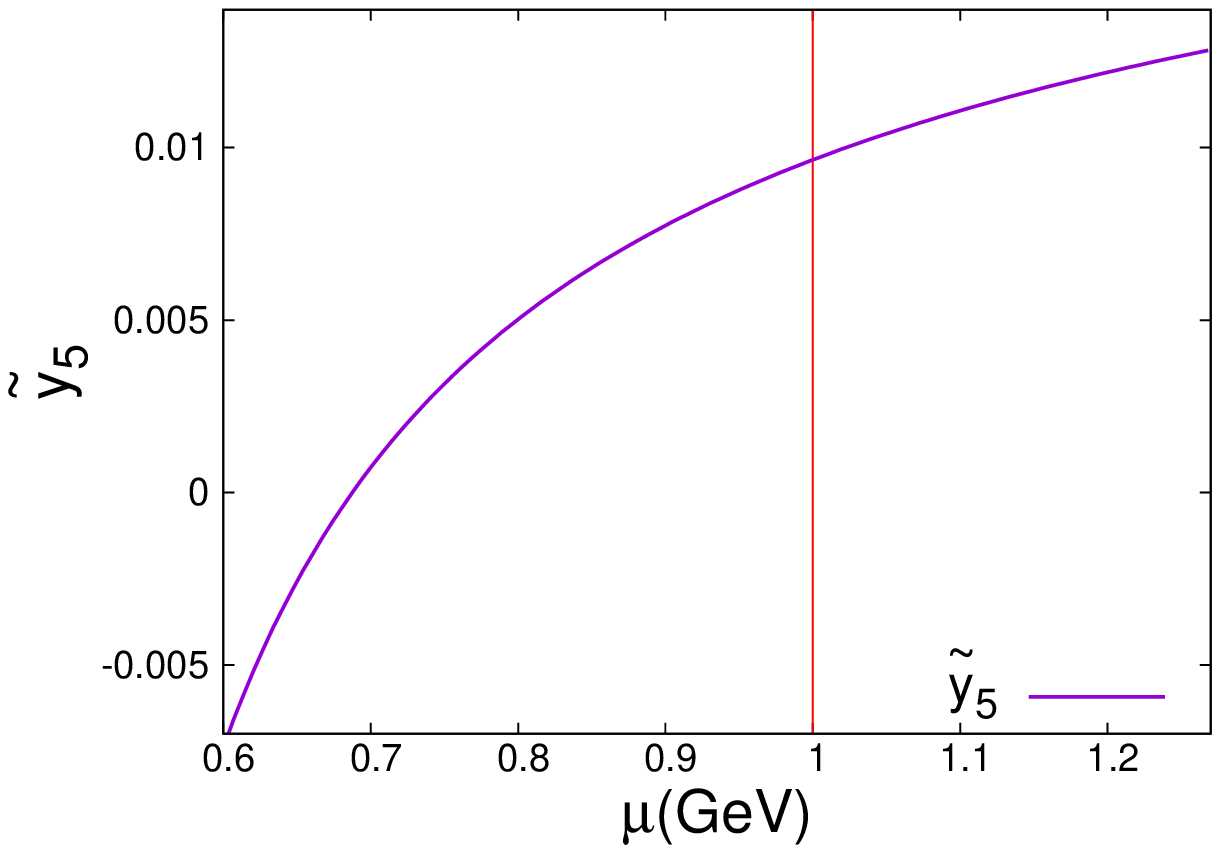}
\end{center}

\end{minipage}
\begin{minipage}{0.5\hsize}
 
\begin{center}
\includegraphics[width=7.5cm]{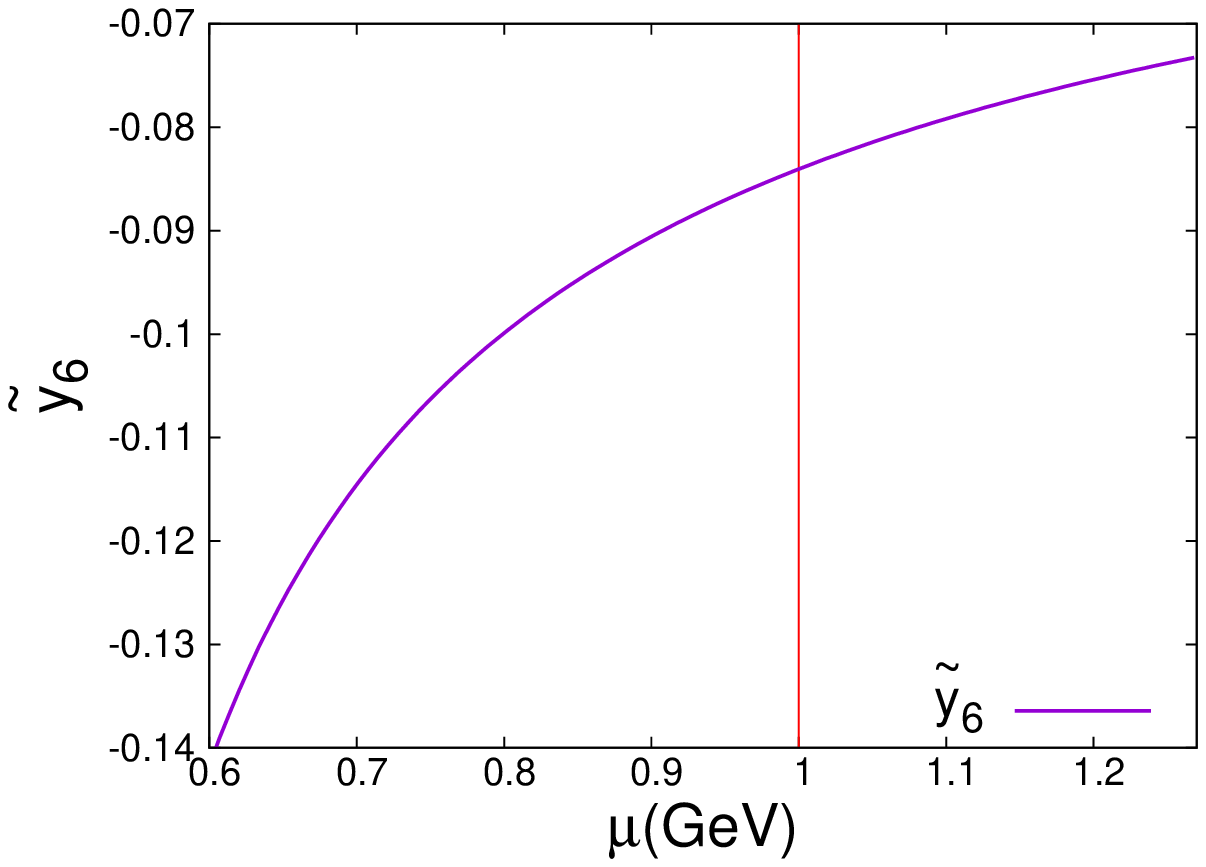}
\end{center}

\end{minipage}
\caption{\label{fig:ytilde}
The behavior of the Wilson coefficients $\tilde y_5$ and $\tilde y_6$ for the scale below the charm quark mass.
The red vertical line denotes the hadronic renormalization scale ($\mu =1$ GeV) we have chosen.
}
\end{figure}

After running numerically the initial Wilson coefficients $\tilde y_i$ and $\tilde z_i$ down to $\mu = 1$ GeV, we obtain
\begin{equation}
\tilde {\bf z} (\mu = 1 \, {\rm GeV})
=
\left(
\begin{array}{c}
-0.457 \cr
1.23 \cr
9.86 \times 10^{-3} \cr
-2.60 \times 10^{-2}  \cr
1.07 \times 10^{-2}  \cr
-1.81 \times 10^{-2} \cr
\end{array}
\right)
,
\ \ \ \ \ \ \ \ 
\tilde {\bf y} (\mu = 1 \, {\rm GeV})
=
\left(
\begin{array}{c}
0 \cr
0 \cr
2.76 \times 10^{-2} \cr
-5.16 \times 10^{-2} \cr
9.65 \times 10^{-3} \cr
-8.40 \times 10^{-2} \cr
\end{array}
\right)
.
\label{eq:zytilde}
\end{equation}
The behavior of the most important coefficients $\tilde z_1$, $\tilde z_2$, $\tilde y_5$, and $\tilde y_6$ are shown in Figs. \ref{fig:ztilde} and \ref{fig:ytilde}.

We must not forget to multiply the operator correction matrix to these arrays to cancel the renormalization scheme dependence.
The correction matrix is given in Appendix \ref{sec:penguinrge}.
After the correction, we get
\begin{equation}
{\bf z} (\mu = 1 \, {\rm GeV})
=
\left(
\begin{array}{c}
-0.107 \cr
1.02 \cr
1.76 \times 10^{-5} \cr
-1.39 \times 10^{-2}  \cr
6.37 \times 10^{-3} \cr
-3.45 \times 10^{-3} \cr
\end{array}
\right)
,
\ \ \ \ \ \ \ \ 
{\bf y} (\mu = 1 \, {\rm GeV})
=
\left(
\begin{array}{c}
0 \cr
0 \cr
1.48 \times 10^{-2} \cr
-4.81 \times 10^{-2} \cr
3.25 \times 10^{-3} \cr
-5.68 \times 10^{-2} \cr
\end{array}
\right)
.
\label{eq:zy}
\end{equation}
Note that the modification due to the correction matrix is not small at the hadronic scale.
In this work, it is important to know the values of both arrays of Wilson coefficients $z_i$ and $y_i$, because we finally have to take their product to form the Jarlskog invariant \cite{jarlskog}, the leading constant of the CP violation of the CKM matrix.
The values of $z_2$ and $y_6$ will be of particular interest in our analysis, since they give the leading contribution to the relevant hadron level effective interaction.
We see that through the renormalization from the weak scale $\mu = m_W$ down to the hadronic scale $\mu = 1 $ GeV, $\tilde y_6$ is enhanced by about 40 times.
The enhancement of the Wilson coefficient of $Q_6$ is due to the next-to-leading logarithmic contribution \cite{buras}. 
This fact signalizes that the analysis of the renormalization group evolution is indispensable in the evaluation of the SM CP violation at low energy.

\subsection{P, CP-odd nuclear force}

Now that we have the $|\Delta S|=1$ effective four-quark interaction at the hadronic scale ($\mu = 1$ GeV), we have to construct the hadron level effective theory and and derive the CP-odd nuclear force.
It is implicitly assumed that the hadron level effective theory below the hadronic scale does not run much.
As seen previously, the leading CP violation of the CKM matrix manifests itself through the Jarlskog invariant.
We therefore have to combine the $|\Delta S| = 1$ hadron level effective interactions which issue from the two terms of Eq. (\ref{eq:effhamibelowmc}) to obtain CP-violating processes.

\begin{figure}[htb]
\begin{center}
\includegraphics[width=12cm]{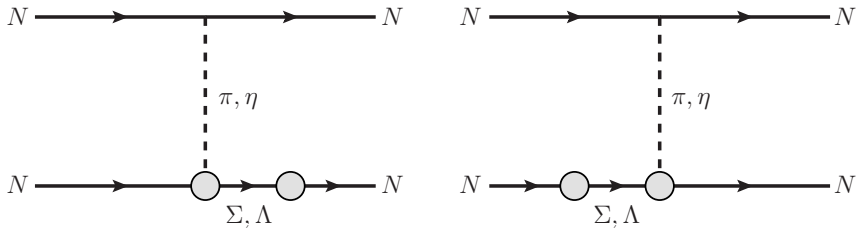}
\caption{\label{fig:CP-odd_piNN}
The CP-odd nuclear force generated by the one-pion (or $\eta$ meson) exchange $|\Delta S| =1$ processes.
The grey blobs denote the hadron level weak vertices.
}
\end{center}
\end{figure}

\begin{figure}[htb]
\begin{center}
\includegraphics[width=12cm]{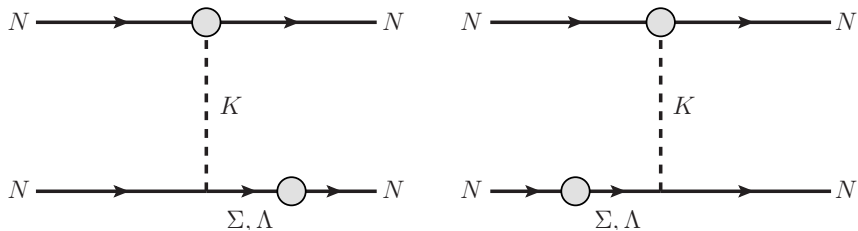}
\caption{\label{fig:CP-odd_KNN}
The CP-odd nuclear force generated by the one-kaon exchange $|\Delta S| =1$ processes.
The grey blobs denote the hadron level weak vertices.
}
\end{center}
\end{figure}

In considering the CP-odd nuclear force, we neglect the contribution from vector meson exchange, since mesons heavier than the $\rho$ meson have small effect on the nuclear EDM \cite{liu,stetcu}.
Moreover, we neglect the appearance of excitations of hyperons, for which the energy required to obtain them is much higher than that of the octet hyperons.
Under those conditions, the only allowed leading order processes to generate the CP-odd nuclear force are the one-pion, one-eta meson, and one-kaon exchange $N-N$ interactions, shown in Figs. \ref{fig:CP-odd_piNN} and  \ref{fig:CP-odd_KNN}, respectively.
There is also an additional diagram with a transition from $\pi$ (or $\eta$) to $K$ meson, but this contribution is known to be small \cite{smCPVNN}.
The CP-odd nuclear force is thus a combination of $|\Delta S| = 1$ effective meson-baryon interactions and $|\Delta S| = 1$ hyperon-nucleon transitions.
Those hadron level effective interactions are calculated in the factorization approach.

\subsubsection{$|\Delta S| =  1$ meson-baryon vertex in the factorization approximation}

We calculate the hadron matrix elements of the $|\Delta S| = 1$ meson-baryon vertex.
They are obtained using the factorization approach, with the vacuum saturation approximation and the partially conserved axial vector current (PCAC) assertions.

In the factorization approach, the $|\Delta S| = 1$ four-quark operators $Q_1$, $Q_2$, $Q_3$ and $Q_4$ generate vector meson-baryon vertices or light pseudoscalar meson vertices with a factor of light quark mass. 
Their effect is therefore small.
Sizable $|\Delta S| = 1$ meson-baryon interactions have to be generated by the right-handed penguin operators $Q_5$ and $Q_6$ so as to obtain a pseudoscalar quark bilinear which will become the interpolating field of a light pseudoscalar meson.
After Fierz transformation, $Q_5$ and $Q_6$ become
\begin{eqnarray}
Q_5
&=&
\frac{2}{3} \sum_{q=u,d,s} \bar s (1+\gamma_5) q \, \bar q (1-\gamma_5) d 
+ 4 \sum_{q=u,d,s} \sum_a \bar s (1+\gamma_5) t_a q \, \bar q (1-\gamma_5) t_a d 
,
\label{eq:q5fierz}
\\
Q_6
&=&
2 \sum_{q=u,d,s} \bar s (1+\gamma_5) q \, \bar q (1-\gamma_5) d 
,
\label{eq:q6fierz}
\end{eqnarray}
where $t_a$ is the generator of the color $SU(3)_c$ group.
The second term of $Q_5$ does not contribute to the meson-baryon vertex, since the color octet asymptotic states cannot exist due to the color confinement.

To obtain meson-baryon vertices in the factorization approach, one of the two quark bilinears of Eqs. (\ref{eq:q5fierz}) and (\ref{eq:q6fierz}) must be considered as an interpolating operator of the pseudoscalar meson.
Therefore, at least one of the quark bilinears must have a pseudoscalar bilinear $\bar q \gamma_5 q'$.
The contribution from the four-quark operators with two pseudoscalar quark bilinears to the meson-baryon vertex is canceled by 
additional effect involving the chiral condensate, owing to the chiral symmetry \cite{donoghue}.
The $|\Delta S| = 1$ meson-baryon vertices are thus generated by the four-quark operators involving only one pseudoscalar bilinear, like $\bar s \gamma_5 q \cdot \bar q d$ and $\bar s q \cdot \bar q \gamma_5 d$.
By writing explicitly the relevant terms of the effective hamiltonian (\ref{eq:effhamibelowmc}), we have 
\begin{eqnarray}
{\cal H}_{eff}^{P\hspace{-.5em}/\,}
&=&
\frac{G_F}{\sqrt{2}}
\Biggl\{
V_{us}^* V_{ud}
\Biggl[
\frac{2}{3} z_5 + 2 z_6
\Biggr]
-
V_{ts}^* V_{td}
\Biggl[
\frac{2}{3} y_5 + 2 y_6
\Biggr]
\Biggr\}
\nonumber\\
&&\times 
\Bigl\{
\bar s \gamma_5 u \cdot \bar u d
-\bar s u \cdot \bar u \gamma_5 d
+\bar s \gamma_5 d \cdot \bar d d
-\bar s d \cdot \bar d \gamma_5 d
+\bar s \gamma_5 s \cdot \bar s d
-\bar s s \cdot \bar s \gamma_5 d
\Bigr\}
\nonumber\\
&&
+ ({\rm h.c.})
. 
\label{eq:relevanteffhami}
\end{eqnarray}
The terms with the CKM matrix elements $V_{us}^* V_{ud}$ will not be important, since the largest contribution of $V_{us}^* V_{ud}$ are brought by the $Q_2$ operator through the hyperon-nucleon interaction.
In this subsubsection, we neglect terms with $V_{us}^* V_{ud}$.

The factorization of the $|\Delta S|=1$ four-quark operators works as \cite{factorizationedm1,factorizationedm2,factorizationedm3,pvcpvhamiltonian2,factorizationedm4,factorizationedm5,factorizationedm6,yamanakabook,rpvlinearprogramming}
\begin{eqnarray}
\langle m B | \bar s \gamma_5 q \cdot \bar q d | B'  \rangle
&\approx &
\langle m | \bar s \gamma_5 q | 0 \rangle \langle B | \bar q d | B' \rangle
\nonumber\\
&&
+\langle 0 | \bar q d | 0 \rangle
\Biggl[
\langle m B | \bar s \gamma_5 q | B' \rangle
+
\frac{1}{m_{m'}^2}\langle m B | {\cal L}_{\rm QCD} | B' m' \rangle \langle m'| \bar s \gamma_5 q | 0 \rangle
\Biggr]
\nonumber\\
&\approx &
\langle m | \bar s \gamma_5 q | 0 \rangle \langle B | \bar q d | B' \rangle
,
\end{eqnarray}
with the pseudoscalar mesons $m$, $m'$ and baryons $B$, $B'$.
Terms of the second line cancel in the flavor $SU(3)$ for light quarks \cite{factorizationedm6}.
The factorization method with the vacuum saturation approximation was first used in the calculation of the strangeness changing  $K^0 - \bar K^0$ mixing \cite{gaillard,vacuumsaturation}.
The theoretical uncertainty due to the use of vacuum saturation hypothesis should be $O(100\%)$ \cite{factorizationedm1,factorizationedm2}.

The hadron level effective $|\Delta S| =1 $ meson-baryon interaction we want to calculate is
\begin{eqnarray}
{\cal L}_{mBB'}^{P\hspace{-.5em}/\,}
&=&
\bar g_{K^- pn} K^+ \bar p n
+
\bar g_{\bar K^0 p p} K^0 \bar p p
+
\bar g_{\bar K^0 n n} K^0 \bar n n
\nonumber\\
&&
+
\bar g_{\pi^+ \Sigma^0 p} \pi^- \bar \Sigma^0 p
+
\bar g_{\pi^0 \Sigma^+ p} \pi^0 \bar \Sigma^+ p
+
\bar g_{\pi^0 \Sigma^0 n} \pi^0 \bar \Sigma^0 n
+
\bar g_{\pi^+ \Lambda p} \pi^- \bar \Lambda p
+
\bar g_{\pi^0 \Lambda n} \pi^0 \bar \Lambda n
\nonumber\\
&&
+
\bar g_{\eta \Sigma^+ p} \eta \bar \Sigma^+ p
+
\bar g_{\eta \Sigma^0 n} \eta \bar \Sigma^0 n
+
\bar g_{\eta \Lambda n} \eta \bar \Lambda n
\nonumber\\
&&
+
({\rm h.c.})
.
\label{eq:cpvmeson-baryon}
\end{eqnarray}
It is to be noted that the above interaction is parity violating.
The coupling constants are expressed in terms of matrix elements to which the vacuum saturation approximation was applied:
\begin{eqnarray}
\bar g_{K^- p n}
&=&
G_y 
\langle K^- p | \bar s \gamma_5 u \cdot  \bar u d | n \rangle
\approx
G_y 
\langle K^-  | \bar s \gamma_5 u | 0 \rangle \langle p |  \bar u d | n \rangle
,
\\
\bar g_{\bar K^0 p p}
&=&
G_y 
\langle \bar K^0 p| \bar s \gamma_5 d \cdot  \bar d d |p \rangle
\approx
G_y 
\langle \bar K^0 | \bar s \gamma_5 d | 0 \rangle \langle p |  \bar d d |p \rangle
,
\\
\bar g_{\bar K^0 n n}
&=&
G_y 
\langle \bar K^0 n | \bar s \gamma_5 d \cdot  \bar d d | n \rangle
\approx
G_y 
\langle \bar K^0  | \bar s \gamma_5 d | 0 \rangle \langle n |  \bar d d | n \rangle
,
\\
\bar g_{\pi^+ \Sigma^0 p}
&=&
-G_y 
\langle \pi^+ \Sigma^0 | \bar s u \cdot  \bar u \gamma_5 d | p \rangle
\approx
-G_y 
\langle \pi^+  | \bar u \gamma_5 d | 0 \rangle \langle \Sigma^0 | \bar s u | p \rangle
,
\\
\bar g_{\pi^0 \Sigma^+ p}
&=&
-G_y 
\langle \pi^0 \Sigma^+ | \bar s d \cdot  \bar d \gamma_5 d | p \rangle
\approx
-G_y 
\langle \pi^0  | \bar d \gamma_5 d | 0 \rangle \langle \Sigma^+ |  \bar s d | p \rangle
,
\\
\bar g_{\pi^0 \Sigma^0 n}
&=&
-G_y 
\langle \pi^0 \Sigma^0 | \bar s d \cdot  \bar d \gamma_5 d | n \rangle
\approx
-G_y 
\langle \pi^0 | \bar d \gamma_5 d | 0 \rangle \langle \Sigma^0 | \bar s d | n \rangle
,
\\
\bar g_{\pi^+ \Lambda p}
&=&
-G_y 
\langle \pi^+ \Lambda | \bar s u \cdot  \bar u \gamma_5 d | p \rangle
\approx
-G_y 
\langle \pi^+ | \bar u \gamma_5 d | 0 \rangle \langle \Lambda | \bar s u | p \rangle
,
\\
\bar g_{\pi^0 \Lambda n}
&=&
-G_y 
\langle \pi^0 \Lambda | \bar s d \cdot  \bar d \gamma_5 d | n \rangle
\approx
-G_y 
\langle \pi^0 | \bar d \gamma_5 d | 0 \rangle \langle \Lambda | \bar s d | n \rangle
,
\\
\bar g_{\eta \Sigma^+ p}
&=&
G_y 
\langle \eta \Sigma^+ | \bar s d \cdot  (\bar s \gamma_5 s - \bar d \gamma_5 d) | p \rangle
\approx
G_y 
\langle \eta  | \bar s \gamma_5 s - \bar d \gamma_5 d | 0 \rangle \langle \Sigma^+ |  \bar s d | p \rangle
,
\\
\bar g_{\eta \Sigma^0 n}
&=&
G_y 
\langle \eta \Sigma^0 | \bar s d \cdot   (\bar s \gamma_5 s - \bar d \gamma_5 d) | n \rangle
\approx
G_y 
\langle \eta | \bar s \gamma_5 s - \bar d \gamma_5 d | 0 \rangle \langle \Sigma^0 | \bar s d | n \rangle
,
\\
\bar g_{\eta \Lambda n}
&=&
G_y 
\langle \eta \Lambda | \bar s d \cdot  (\bar s \gamma_5 s - \bar d \gamma_5 d) | n \rangle
\approx
G_y 
\langle \eta | \bar s \gamma_5 s - \bar d \gamma_5 d | 0 \rangle \langle \Lambda | \bar s d | n \rangle
,
\end{eqnarray}
where $G_y  = V_{ts}^* V_{td} \frac{G_F}{\sqrt{2}} \bigl[  \frac{2}{3} y_5 (\mu = 1 \, {\rm GeV}) + 2 y_6 (\mu = 1 \, {\rm GeV}) \bigr]$ [see Eq. (\ref{eq:zy})].
From the above formulae of the CP-odd couplings, we can have a rough idea that the terms of the four-quark interaction hamiltonian (\ref{eq:relevanteffhami}) with $\bar s \gamma_5 d$ contribute to the CP-odd kaon-exchange nuclear force, and those with $\bar s d$ to the CP-odd pion-exchange (or $\eta$-exchange) nuclear force.
In this sense, it is very important to note that the sign of the terms with $\bar s \gamma_5 d$ and those with $\bar s d$ are opposite in Eq. (\ref{eq:relevanteffhami}), since this fact controls the relative sign between the contributions from the CP-odd pion (or $\eta$) and kaon exchange nuclear forces.

The pseudoscalar matrix element $\langle m | \bar q' \gamma_5 q | 0 \rangle$ appearing after applying the vacuum saturation approximation can be transformed to the chiral condensate using the PCAC reduction
\begin{equation}
\langle  \pi^c \beta | \, O_i \, | \, \alpha \rangle \approx \frac{i}{f_\pi } \langle \beta  | \, [ O_i, Q_5^c ]\,  | \, \alpha \rangle \ ,
\end{equation}
where $Q_5^c \equiv \int d {\bf x} \, (J_5^c)_0 (x)$ is the axial charge.
The PCAC reductions of the required pseudoscalar matrix elements are collected in Appendix \ref{sec:pcac}.

The baryon scalar densities $\langle B | \bar q \Gamma' q'' | N \rangle$ cannot be transformed further, and are input parameters in this discussion.
For the kaon creating/annihilating process, the nucleon matrix elements $\langle p | \bar d d  | p \rangle$, $\langle n | \bar d d  | n \rangle$, and $\langle p | \bar u d  | n \rangle$ are needed.
Those are calculable from the data of the pion-nucleon sigma term and the proton-neutron mass difference.
From the phenomenology, the pion-nucleon sigma term is given by \cite{cheng,gasser1,gasser2,gasser3,borasoy,alarcon1,alarcon2,ren}
\begin{equation}
\sigma_{\pi N} =
\frac{m_u + m_d}{2}\langle N | \bar uu + \bar dd | N \rangle
=
45 \, {\rm MeV}
.
\end{equation}
The results of lattice QCD calculations are also in agreement with this value \cite{weise1,weise2,jlqcd1,qcdsf1,young1,qcdsf2,durr1,dinter,bhattacharya1,etm2,etm3,durr2,yang}.
The up and down quark masses at the hadronic scale $\mu =1$ GeV are $m_u = 2.9$ MeV and $m_d = 6.0$ MeV, respectively.
Those masses were obtained through the two-loop level renormalization group evolution \cite{tarasov,gorishnii}, with the initial conditions $m_u = 2.3$ MeV and $m_d = 4.8$ MeV at the scale $\mu = 2 $ GeV (Particle Data Group) \cite{pdg}.
The isoscalar nucleon scalar density is then given by
\begin{equation}
\langle N | \bar uu + \bar dd | N \rangle
=
10
,
\label{eq:nucleonscalardensity}
\end{equation}
at $\mu =1$ GeV.
Note that this value is quite large compared with other baryon scalar matrix elements, which are introduced below.

To obtain $\langle p | \bar d d  | p \rangle$, $\langle n | \bar d d  | n \rangle$, and $\langle p | \bar u d  | n \rangle$, the information of the isovector nucleon matrix element is needed.
The isovector nucleon scalar matrix element is given by \cite{rpvbetadecay1,rpvedm,rpvbetadecay2,gonzales-alonso,yamanakabook}
\begin{equation}
\langle p | \bar u d  | n \rangle
=
\langle p | \bar uu - \bar dd | p \rangle
=
\frac{m_n^{(0)}-m_p^{(0)} }{m_d -m_u}
,
\end{equation}
where $m_n^{(0)}$ and $m_p^{(0)}$ are the proton and neutron masses without electromagnetic contribution.
The first equality follows from isospin symmetry.
The electromagnetic contribution has been calculated in Ref. \cite{thomas}, yielding $\delta M_{p-n}^\gamma = 1.04 \pm 0.11$ MeV.
The proton-neutron mass difference without electromagnetic effect is thus $m_n^{(0)}-m_p^{(0)} = 1.2933322 \pm 0.0000004 + 1.04 \pm 0.11 = 2.33 \pm 0.11$ MeV, where the first term is the physical proton-neutron mass splitting \cite{pdg,mohr}.
At the hadronic scale $\mu = 1$ GeV, the isovector nucleon scalar matrix elements are then
\begin{equation}
\langle p | \bar u d  | n \rangle
=
\langle p | \bar uu - \bar dd | p \rangle
=
0.74
,
\end{equation}
where we have again used the renormalized up and down quark masses at $\mu = 1$ GeV.
Here the isospin symmetry was assumed.
The strange content of the nucleon $\langle N | \bar ss | N \rangle$ are known to be smaller than the light quark contribution from lattice QCD calculations \cite{jlqcd2,jlqcd3,qcdsf1,young1,durr1,dinter,milc1,qcdsf2,engelhardt,gong,etm1,etm2,junnarkar,yang} and from phenomenology \cite{alarcon3,gubler}, so we neglect it.

For the pion and eta creating/annihilating processes, the values of the hyperon scalar transition matrix elements $\langle \Lambda | \bar s u  | p \rangle$, $\langle \Lambda | \bar s d  | n \rangle$, $\langle \Sigma^0 | \bar s u  | p \rangle$, $\langle \Sigma^0 | \bar s d  | n \rangle$, and $\langle \Sigma^+ | \bar s d  | p \rangle$ are required. 
The matrix elements involving $\Lambda$, renormalized at the scale $\mu = 1$ GeV, are given by
\begin{eqnarray}
\langle \Lambda | \bar s u  | p \rangle
&=&
\frac{m_N + m_\Sigma - 2m_\Xi }{\sqrt{6} m_s}
\approx
\sqrt{\frac{3}{2}} \cdot \frac{m_N -m_\Lambda}{m_s}
=
-1.80
,
\\
\langle \Lambda | \bar s d  | n \rangle
&=&
\frac{m_N + m_\Sigma - 2 m_\Xi }{\sqrt{6} m_s}
\approx
\sqrt{\frac{3}{2}} \cdot \frac{m_N -m_\Lambda}{m_s}
=
-1.80
,
\end{eqnarray}
and those involving $\Sigma$'s by
\begin{eqnarray}
\langle \Sigma^0 | \bar s u  | p \rangle
&=&
\frac{m_N - m_\Sigma}{\sqrt{2} m_s}
=
-1.50
,
\\
\langle \Sigma^0 | \bar s d  | n \rangle
&=&
\frac{ m_\Sigma - m_N}{\sqrt{2} m_s}
=
\ \ 1.50
,
\\
\langle \Sigma^+ | \bar s d  | p \rangle
&=&
\frac{m_N - m_\Sigma}{m_s}
=
-2.12
,
\end{eqnarray}
where we have used the isospin symmetry.
The strange quark mass at the hadronic scale $\mu =1$ GeV is $m_s = 120$ MeV.
As for the up and down quark masses, we have again used the two-loop level renormalization group evolution \cite{tarasov,gorishnii}, with the initial value $m_s = 95$ MeV at $\mu = 2 $ GeV \cite{pdg}.
The baryon masses are $m_\Lambda = 1115$ MeV and $m_{\Sigma}=1193$ MeV \cite{pdg}.

The couplings constants of the P-odd meson-baryon interactions
\begin{equation}
{\cal L}_{mBB'}^{P\hspace{-.5em}/\,} 
=
\bar g_{mBB'} m \bar BB'
+ ({\rm h.c.})
,
\end{equation}
are summarized as follows:
\begin{eqnarray}
\bar g_{K^- pn}
&=&
i
G_y 
\frac{1}{\sqrt{2} f_K}
\langle 0 | \bar q q + \bar s s | 0 \rangle
\langle p| \bar u d |n \rangle
,
\\
\bar g_{\bar K^0 p p}
&=&
i
G_y 
\frac{1}{\sqrt{2} f_K}
\langle 0 | \bar q q + \bar s s | 0 \rangle
\langle p| \bar d d |p \rangle
,
\\
\bar g_{\bar K^0 n n}
&=&
i
G_y 
\frac{1}{\sqrt{2} f_K}
\langle 0 | \bar q q + \bar s s | 0 \rangle
\langle n | \bar d d | n \rangle
,
\\
\bar g_{\pi^+ \Sigma^0 p}
&=&
-i
G_y 
\frac{\sqrt{2}}{f_\pi}
\langle 0 | \bar q q | 0 \rangle
\langle \Sigma^0 | \bar s u | p \rangle
,
\\
\bar g_{\pi^0 \Sigma^+ p}
&=&
i
G_y 
\frac{1}{f_\pi}
\langle 0 | \bar q q | 0 \rangle
\langle \Sigma^+ | \bar s d | p \rangle
,
\\
\bar g_{\pi^0 \Sigma^0 n}
&=&
i
G_y 
\frac{1}{f_\pi}
\langle 0 | \bar q q | 0 \rangle
\langle \Sigma^0 | \bar s d | n \rangle
,
\\
\bar g_{\pi^+ \Lambda p}
&=&
-i
G_y 
\frac{\sqrt{2}}{f_\pi}
\langle 0 | \bar q q | 0 \rangle
\langle \Lambda | \bar s u | p \rangle
,
\\
\bar g_{\pi^0 \Lambda n}
&=&
i
G_y 
\frac{1}{f_\pi}
\langle 0 | \bar q q | 0 \rangle
\langle \Lambda | \bar s d | n \rangle
,
\end{eqnarray}
\begin{eqnarray}
\bar g_{\eta \Sigma^+ p}
&=&
-i
G_y 
\frac{1}{\sqrt{3}f_\eta}
\langle 0 | \bar q q + 2 \bar s s | 0 \rangle \langle \Sigma^+ |  \bar s d | p \rangle
,
\\
\bar g_{\eta \Sigma^0 n}
&=&
-i
G_y 
\frac{1}{\sqrt{3}f_\eta}
\langle 0 | \bar q q + 2 \bar s s | 0 \rangle \langle \Sigma^0 | \bar s d | n \rangle
,
\\
\bar g_{\eta \Lambda n}
&=&
-i
G_y 
\frac{1}{\sqrt{3}f_\eta}
\langle 0 | \bar q q + 2 \bar s s | 0 \rangle \langle \Lambda | \bar s d | n \rangle
,
\end{eqnarray}
where $f_\pi = 93$ MeV, $f_K = 1.198 f_\pi $, and $f_\eta \approx f_{\eta_8} = 1.2 f_\pi $ \cite{pdg}.
Here the chiral condensates issue from the PCAC reduction of the pseudoscalar matrix elements $\langle m | \bar q' \gamma_5 q | 0 \rangle$.
The chiral condensate is also a renormalization dependent quantity.
Its explicit expression is given by the Gell-Mann Oakes-Renner relation
\begin{equation}
\langle 0 | \bar q q | 0 \rangle
=
-\frac{m_\pi^2 f_\pi^2}{m_u + m_d}
=
-(265\, {\rm MeV})^3  \ \ \ (\mu = 1\, {\rm GeV})
.
\end{equation}
We see that the quark masses in the denominator bring the scale dependence.
The strange chiral condensate $\langle 0 | \bar s s | 0 \rangle$ has a value comparable to the light quark condensate $\langle 0 | \bar q q | 0 \rangle$.
Recently, the lattice QCD calculation has obtained  \cite{mcneile}
\begin{equation}
\langle 0 | \bar s s | 0 \rangle
= 
(1.08 \pm 0.16 ) \times
\langle 0 | \bar q q | 0 \rangle
,
\end{equation}
which is larger than the previous estimates based on QCD sum rules, predicting ratios smaller than one \cite{narison,dominguez}.
In this work, we assume $\langle 0 | \bar q q | 0 \rangle \approx \langle 0 | \bar s s | 0 \rangle$ which is well enough within the accuracy required in this work.

\subsubsection{Hyperon-nucleon transition}

We now turn to the derivation of the hyperon-nucleon transition in the factorization approach.
The hyperon-nucleon transition was studied in the context of the nonleptonic hyperon decay \cite{feldman,pati,hara,itoh,graham,Guralnik,ahmed,shifman,vainshtein2,donoghue2,donoghue3,henley,tandean,Zenczykowski,hiyamahyperon-nucleon}.
The general form of the hyperon-nucleon transition lagrangian relevant in this work is given by
\begin{equation}
{\cal L}_{YN}
=
a_{p \Sigma^+} \bar p \Sigma^+
+a_{n \Sigma^0} \bar n \Sigma^0
+a_{p \Lambda} \bar n \Lambda
+({\rm h.c.})
.
\label{eq:hyperon-nucleon}
\end{equation}
As we have mentioned in the beginning of the section, we neglect the excited baryon contributions in the intermediate states.
The hyperon-nucleon transition is therefore parity conserving.

\begin{figure}[htb]
\begin{center}
\includegraphics[width=14cm]{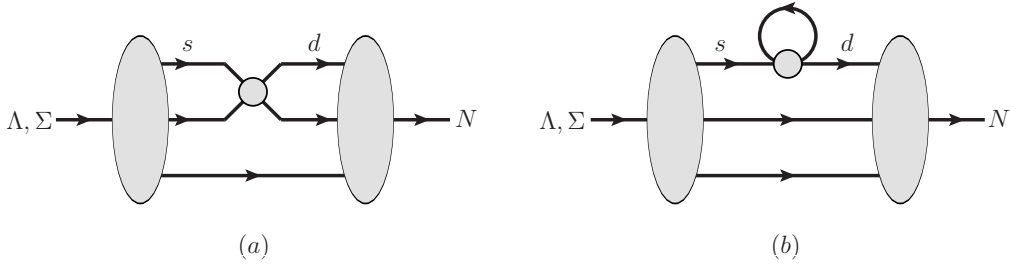}
\caption{\label{fig:hyperon-nucleon}
Schematic picture of the hyperon-nucleon transition from the $\Delta S =1$ four-quark interactions (grey blob) in the factorization approach.
The quark loop of the right figure $(b)$ denotes the quark chiral condensate.
}
\end{center}
\end{figure}

In the factorization approach, the $|\Delta S| =1$ four-quark interactions have two distinct contributions to the hyperon-nucleon transition (see Fig. \ref{fig:hyperon-nucleon}).
The first one is the contribution from the quark-quark interaction [Fig. \ref{fig:hyperon-nucleon} (a)].
This cannot be calculated using the chiral techniques.
In this work, we use the result of Ref. \cite{hiyamahyperon-nucleon}, where the matrix elements of the hyperon-nucleon transition are calculated in the nonrelativistic constituent quark model.
There the nonrelativistically reduced operator $Q_2^{NR} = Q_2 = - Q_1$ was considered\footnote{In Ref. \cite{hiyamahyperon-nucleon}, the definition of the operators $O_1 \equiv Q_2$ and $O_2 \equiv Q_1$ was adopted.}, and the penguin operators were neglected.
The explicit expressions of the transition couplings in terms of the quark-quark interaction are
\begin{eqnarray}
a_{p \Sigma^+} (q-q)
&=&
V_{us} V_{ud}^* \frac{G_F}{\sqrt{2}} (z_1 -z_2)
\langle p | Q_2^{NR} | \Sigma^+ \rangle
,
\\
a_{n \Sigma^0} (q-q) 
&=&
V_{us} V_{ud}^* \frac{G_F}{\sqrt{2}} (z_1 -z_2)
\langle n | Q_2^{NR} | \Sigma^0 \rangle
,
\\
a_{n \Lambda} (q-q)
&=&
V_{us} V_{ud}^* \frac{G_F}{\sqrt{2}} (z_1 -z_2)
\langle n | Q_2^{NR} | \Lambda \rangle
,
\end{eqnarray}
with the hyperon-nucleon transition matrix elements \cite{hiyamahyperon-nucleon}
\begin{eqnarray}
\langle p | Q_2^{NR} | \Sigma^+ \rangle
&=&
2.76 \times 10^{-2} {\rm GeV}^3
,
\\
\langle n | Q_2^{NR} | \Sigma^0 \rangle
&=&
-1.95 \times 10^{-2} {\rm GeV}^3
,
\\
\langle n | Q_2^{NR} | \Lambda \rangle
&=&
-9.65 \times 10^{-3} {\rm GeV}^3
.
\end{eqnarray}
For the second line, we have used the isospin symmetry $\langle n | Q_2^{NR} | \Sigma^0 \rangle = -\frac{1}{\sqrt{2}} \langle p | Q_2^{NR} | \Sigma^+ \rangle$.
The Wilson coefficients $z_1$ and $z_2$ are renormalized at $\mu = 1$ GeV [see Eq. (\ref{eq:zy})].

The second contribution of the $|\Delta S| =1$ four-quark interactions to the hyperon-nucleon transition in the factorization approach is the one-body process [Fig. \ref{fig:hyperon-nucleon} (b)].
This contribution is proportional to the chiral condensate, and are only generated by the right-handed penguin operators $Q_5$ and $Q_6$.
The explicit expression is given as
\begin{eqnarray}
a_{p \Sigma^+} (1q)
&=&
-V_{us} V_{ud}^* \frac{G_F}{\sqrt{2}} \Biggl[ \frac{2}{3} z_5 +2 z_6 \Biggr]
\langle p | \bar s d | \Sigma^+ \rangle
\langle 0 | \bar dd + \bar ss | 0 \rangle
,
\\
a_{n \Sigma^0} (1q) 
&=&
-V_{us} V_{ud}^* \frac{G_F}{\sqrt{2}} \Biggl[ \frac{2}{3} z_5 +2 z_6 \Biggr]
\langle n | \bar s d | \Sigma^0 \rangle
\langle 0 | \bar dd + \bar ss | 0 \rangle
,
\\
a_{n \Lambda} (1q)
&=&
-V_{us} V_{ud}^* \frac{G_F}{\sqrt{2}} \Biggl[ \frac{2}{3} z_5 +2 z_6 \Biggr]
\langle n | \bar s d | \Sigma^0 \rangle
\langle 0 | \bar dd + \bar ss | 0 \rangle
.
\end{eqnarray}
As the Wilson coefficients of the right-handed penguin operators ($Q_5$, and $Q_6$) are much smaller than that of $Q_2$ [see Eq. (\ref{eq:zy})],  the one-body process is subleading.

We must also note that the hyperon-nucleon transition receives contribution from the Wilson coefficients $y_i$ ($i=3,4,5,6$) (coefficients of $V_{ts}^* V_{td}$).
In this case, the hyperon-nucleon transition matrix elements for the penguin operators $Q_3$, $Q_4$, $Q_5$, and $Q_6$ are needed, but those are so far unknown.
This effect is however subleading, since they do not involve the largest Wilson coefficient $z_2$ in the P-odd $|\Delta S|=1$ meson-baryon vertex.
In this work, we therefore neglect it.

\subsubsection{CP-odd nuclear force with pion-, $\eta$- and kaon-exchange}

Now that we have the expressions for the meson-baryon interaction and the hyperon-nucleon transition, we can combine them to form the CP-odd nuclear force.
By combining the $|\Delta S|=1$ meson-baryon interactions with pion fields of Eq. (\ref{eq:cpvmeson-baryon}) and the hyperon-nucleon transition lagrangian of Eq. (\ref{eq:hyperon-nucleon}), we obtain the following CP-odd pion-nucleon interaction in the charge base
\begin{equation}
{\cal L}_{\pi NN}
=
\bar g_{\pi^0 p p} \pi^0 \bar p p  
+\bar g_{\pi^0 n n} \pi^0 \bar n n  
+\bar g_{\pi^+ n p} \pi^- \bar n p
+\bar g_{\pi^+ n p}^* \pi^+ \bar p n
,
\end{equation}
with the couplings
\begin{eqnarray}
\bar g_{\pi^0 p p}
&=& 
- \frac{2 \, {\rm Re} ( \bar g_{\pi^0 \Sigma^+ p} a_{p \Sigma^+} ) }{m_N -m_\Sigma} 
\nonumber\\
&=&
\frac{2 J G_{yz} \langle 0 | \bar q q | 0 \rangle
\langle \Sigma^+ | \bar s d | p \rangle
\langle p | Q_2^{NR} | \Sigma^+ \rangle
 }{f_\pi (m_N -m_\Sigma )} 
,
\\
\bar g_{\pi^0 n n}
&=&
- \frac{2 \, {\rm Re} ( \bar g_{\pi^0 \Lambda n} a_{n \Lambda} ) }{m_N -m_\Lambda} 
- \frac{2 \, {\rm Re} ( \bar g_{\pi^0 \Sigma^0 n} a_{n \Sigma^0} ) }{m_N -m_\Sigma} 
\nonumber\\
&=&
\frac{2 J G_{yz} \langle 0 | \bar q q | 0 \rangle
 }{f_\pi } 
\Biggl[
\frac{\langle \Lambda | \bar s d | n \rangle
\langle n | Q_2^{NR} | \Lambda \rangle
}{m_N -m_\Lambda }
+
\frac{\langle \Sigma^0 | \bar s d | n \rangle
\langle n | Q_2^{NR} | \Sigma^0 \rangle
}{m_N -m_\Sigma }
\Biggr]
,
\\
{\rm Re} ( \bar g_{\pi^+ np} )
&=&
- \frac{ {\rm Re} ( \bar g_{\pi^+ \Lambda p} a_{n \Lambda} ) }{m_N -m_\Lambda} 
- \frac{ {\rm Re} ( \bar g_{\pi^+ \Sigma^0 p} a_{n \Sigma^0} ) }{m_N -m_\Sigma} 
\nonumber\\
&=&
-\frac{\sqrt{2} J G_{yz} \langle 0 | \bar q q | 0 \rangle
 }{f_\pi } 
\Biggl[
\frac{\langle \Lambda | \bar s u | p \rangle
\langle n | Q_2^{NR} | \Lambda \rangle
}{m_N -m_\Lambda }
+
\frac{\langle \Sigma^0 | \bar s u | p \rangle
\langle n | Q_2^{NR} | \Sigma^0 \rangle
}{m_N -m_\Sigma }
\Biggr]
, \ \ \ \ 
\end{eqnarray}
where $G_{yz} \equiv \frac{G_F^2}{2} (z_1 -z_2) \bigl( \frac{2}{3} y_5 + 2y_6 \bigr)$ with $z_1$, $z_2$, $y_5$, and $y_6$ renormalized at $\mu = 1$ GeV, and $J$ is the CP violating Jarlskog invariant $J = {\rm Im}[V_{us} V_{td} V^*_{ud} V^*_{ts}]= (3.06^{+0.21}_{-0.20} ) \times 10^{-5}$ \cite{jarlskog,pdg}.
For the charged pion-nucleon interaction $\bar g_{\pi^+ np} $, we have only considered the real part which contributes to the CP-violation.
The CP-odd pion-nucleon interaction is often expressed in the isospin base, as \cite{pospelovreview}
\begin{equation}
{\cal L}_{\pi NN} 
=
\sum _{N=p,n} \Big[ \bar g_{\pi NN}^{(0)} \bar N \tau^a N \pi^a 
+ \bar g_{\pi NN}^{(1)} \bar N N \pi^0
+ {\bar g}^{(2)}_{\pi NN}\left (
\Bar N \tau ^{a} N \pi ^{a}-3\bar N \tau ^{z} N \pi ^{0}
\right )
\Big] \, ,
\label{eq:pcpvpinnint}
\end{equation}
where the implicit sum over the isospin index $a$ was taken.
It is therefore practical to convert the CP-odd nuclear force from the charge base to the isospin base.
This is achieved just by changing the basis as
\begin{equation}
\left(
\begin{array}{c}
\bar g^{(0)}_{\pi NN} \cr
\bar g^{(1)}_{\pi NN} \cr
\bar g^{(2)}_{\pi NN} \cr
\end{array}
\right)
=
A_\pi^{-1}
\left(
\begin{array}{c}
\bar g_{\pi^0 pp} \cr
\bar g_{\pi^0 nn} \cr
{\rm Re} ( \bar g_{\pi^+ np} ) \cr
\end{array}
\right)
,
\end{equation}
where we have to take the inverse of the matrix $A_\pi$ given by
\begin{equation}
A_\pi=
\left(
\begin{array}{rrr}
1 & \ \ 1  & -2\ \cr
-1 & 1 & 2\ \cr
\sqrt{2} & 0 & \sqrt{2} \ \cr
\end{array}
\right)
.
\end{equation}
We show here the explicit value of the CP-odd pion-nucleon couplings in the SM:
\begin{eqnarray}
\bar g^{(0)}_{\pi NN}
&=&
-1.1 \times 10^{-17}
,
\\
\bar g^{(1)}_{\pi NN} 
&=&
-1.3 \times 10^{-17}
,
\\
\bar g^{(2)}_{\pi NN}
&=&
3.3 \times 10^{-21}
.
\end{eqnarray}

We now give the one-pion exchange CP-odd nuclear force.
It has been studied and used in many previous works \cite{korkin,liu,stetcu,song,bsaisou2,yamanakanuclearedm}.
The CP-odd one-pion exchange nuclear force in the coordinate representation is \cite{pvcpvhamiltonian1,pvcpvhamiltonian2,pvcpvhamiltonian3,liu}
\begin{eqnarray}
H_{P\hspace{-.5em}/\, T\hspace{-.5em}/\, }^\pi
& = &
 \frac{1}{2m_N} \bigg\{ 
\bar{G}_{\pi}^{(0)}\,{\mathbf \tau}_{1}\cdot {\mathbf \tau}_{2}\, {\mathbf \sigma}_{-}\cdot {\mathbf \nabla} {\cal Y}_{\pi}(r)
\nonumber \\
&& \hspace{3em} 
+\frac{1}{2} \bar{G}_{\pi}^{(1)}\,
( \tau_{+}^{z}\, {\mathbf \sigma}_{-} +\tau_{-}^{z}\,{\mathbf \sigma}_{+} )
\cdot {\mathbf \nabla} {\cal Y}_{\pi}(r)
\nonumber \\
&& \hspace{3em}
+\bar{G}_{\pi}^{(2)}\, (3\tau_{1}^{z}\tau_{2}^{z}- {\mathbf \tau}_{1}\cdot {\mathbf \tau}_{2})\,{\mathbf \sigma}_{-}\cdot {\mathbf \nabla}
{\cal Y}_{\pi}(r)
\bigg\}
,
\label{eq:CPVhamiltonian}
\end{eqnarray}
where we denote the Yukawa function by ${\cal Y}_\pi (r) =  \frac{e^{-m_\pi r }}{4 \pi r}$, with the pion mass $m_\pi = 138$ MeV.
The spin and isospin notations are ${\mathbf \sigma}_{-} \equiv {\mathbf \sigma}_1 -{\mathbf \sigma}_2$, ${\mathbf \sigma}_{+} \equiv {\mathbf \sigma}_1 + {\mathbf \sigma}_2$, ${\mathbf \tau}_{-} \equiv {\mathbf \tau}_1 -{\mathbf \tau}_2$, and ${\mathbf \tau}_{+} \equiv {\mathbf \tau}_1 + {\mathbf \tau}_2$.
The coupling constants of the CP-odd nuclear force with their explicit values are given by
\begin{eqnarray}
\bar G^{(0)}_\pi 
&\equiv& 
-g_{\pi NN} \bar g^{(0)}_{\pi NN}
=
1.6 \times 10^{-16}
,
\label{eq:g0pi}
\\
\bar G^{(1)}_\pi
&\equiv& 
-g_{\pi NN} \bar g^{(1)}_{\pi NN}
=
1.8 \times 10^{-16}
,
\label{eq:g1pi}
\\
\bar G^{(2)}_\pi
&\equiv& 
\ \  g_{\pi NN} \bar g^{(2)}_{\pi NN}
=
4.7 \times 10^{-20}
.
\label{eq:g2pi}
\end{eqnarray}
Note the sign change for the isoscalar and isovector couplings due to the difference of conventions \cite{pospelovreview, korkin} 
[see the minus sign in front of the pion-nucleon coupling in Eq. (\ref{eq:chiral_meson-baryon}) of Appendix \ref{sec:chiral_lagrangian}].
We have adopted $g_{\pi NN} =14.11 \pm 0.20$ for the value of the standard pion-nucleon coupling \cite{ericson}.


The derivation of the eta-exchange CP-odd nuclear force is done in the same way as that for the pion-exchange process.
The following CP-odd eta-nucleon interaction in the charge base is generated
\begin{equation}
{\cal L}_{\eta NN}
=
\bar g_{\eta p p} \eta \bar p p  
+\bar g_{\eta n n} \eta \bar n n  
,
\end{equation}
with the couplings
\begin{eqnarray}
\bar g_{\eta p p}
&=& 
- \frac{2 \, {\rm Re} ( \bar g_{\eta \Sigma^+ p} a_{p \Sigma^+} ) }{m_N -m_\Sigma} 
\nonumber\\
&=&
- \frac{2 J G_{yz} \langle 0 | \bar q q + 2 \bar s s  | 0 \rangle
\langle \Sigma^+ | \bar s d | p \rangle
\langle p | Q_2^{NR} | \Sigma^+ \rangle
 }{\sqrt{3} f_\eta (m_N -m_\Sigma )} 
,
\\
\bar g_{\eta n n}
&=&
- \frac{2 \, {\rm Re} ( \bar g_{\eta \Lambda n} a_{n \Lambda} ) }{m_N -m_\Lambda} 
- \frac{2 \, {\rm Re} ( \bar g_{\eta \Sigma^0 n} a_{n \Sigma^0} ) }{m_N -m_\Sigma} 
\nonumber\\
&=&
- \frac{2 J G_{yz} \langle 0 | \bar q q + 2 \bar s s | 0 \rangle
 }{\sqrt{3} f_\eta } 
\Biggl[
\frac{\langle \Lambda | \bar s d | n \rangle
\langle n | Q_2^{NR} | \Lambda \rangle
}{m_N -m_\Lambda }
+
\frac{\langle \Sigma^0 | \bar s d | n \rangle
\langle n | Q_2^{NR} | \Sigma^0 \rangle
}{m_N -m_\Sigma }
\Biggr]
.
\end{eqnarray}

It is also practical to convert the CP-odd eta-nucleon interaction from the charge base to the isospin base.
The isospin representation is given by
\begin{equation}
{\cal L}_{\eta NN} 
=
\sum _{N=p,n} \Big[ \bar g_{\eta NN}^{(0)} \bar N N \eta
+ \bar g_{\eta NN}^{(1)} \bar N \tau^z N \eta
\Big] \, .
\label{eq:pcpvetannint}
\end{equation}
This is achieved just by changing the basis as
\begin{equation}
\left(
\begin{array}{c}
\bar g^{(0)}_{\eta NN} \cr
\bar g^{(1)}_{\eta NN} \cr
\end{array}
\right)
=
A_\eta^{-1}
\left(
\begin{array}{c}
\bar g_{\eta^0 pp} \cr
\bar g_{\eta^0 nn} \cr
\end{array}
\right)
,
\end{equation}
where we have to take the inverse of the matrix $A_\eta$ given by
\begin{equation}
A_\eta =
\left(
\begin{array}{rr}
1 &  1  \cr
1 & -1  \cr
\end{array}
\right)
.
\end{equation}
The explicit values of the CP-odd eta-nucleon couplings in the SM are
\begin{eqnarray}
\bar g^{(0)}_{\eta NN}
&=&
1.8 \times 10^{-17}
,
\\
\bar g^{(1)}_{\eta NN} 
&=&
1.6 \times 10^{-17}
.
\end{eqnarray}

In previous works, the CP-odd one-$\eta$ exchange nuclear force was often defined as \cite{korkin,liu,stetcu,song,bsaisou2,yamanakanuclearedm}
\begin{eqnarray}
H_{P\hspace{-.5em}/\, T\hspace{-.5em}/\, }^\eta
& = &
 \frac{1}{2m_N} \bigg\{ 
\ \ \ \ \bar{G}_{\eta}^{(0')}\,{\mathbf \sigma}_{-}\cdot {\mathbf \nabla} {\cal Y}_\eta (r)
\nonumber \\
&& \hspace{3em} 
+\frac{1}{2} \bar{G}_\eta^{(1')}\,
( \tau_{-}^{z}\,{\mathbf \sigma}_{+} - \tau_{+}^{z}\, {\mathbf \sigma}_{-}  )
\cdot {\mathbf \nabla} {\cal Y}_\eta (r)
\bigg\}
,
\end{eqnarray}
where we have used the same notation as Eq. (\ref{eq:CPVhamiltonian}).
The mass parameter of the Yukawa function ${\cal Y}_\eta (r)$ has to be replaced by the $\eta$ mass $m_\eta = 550$ MeV.

The coupling constants of the CP-odd nuclear force with their explicit values are given by
\begin{eqnarray}
\bar G^{(0')}_\eta
&\equiv& 
-g_{\eta NN} \bar g^{(0)}_{\eta NN}
=
-4.1 \times 10^{-17}
,
\label{eq:g0eta}
\\
\bar G^{(1')}_\eta
&\equiv& 
-g_{\eta NN} \bar g^{(1)}_{\eta NN}
=
-3.6 \times 10^{-17}
.
\label{eq:g1eta}
\end{eqnarray}
Here we use the phenomenological eta-nucleon coupling constant $g_{\eta NN} = 2.24$ \cite{tiator}, which is not far from the on-shell approximation of the chiral interaction $g_{\eta NN} = \frac{m_N}{\sqrt{3}f_\pi} (3F-D) \approx 3.0$ (see Appendix \ref{sec:chiral_lagrangian} for derivation).
Here we also note the sign change due to difference of conventions \cite{pospelovreview, korkin}.


We now derive the $K$ meson exchange contribution to the P, CP-odd nuclear force (see Fig. \ref{fig:CP-odd_KNN}).
To calculate it, we need the values of the coupling constants of the P, CP-even meson-baryon interactions.
In this work, we take the on-shell approximation of the leading terms of the chiral lagrangian:
\begin{equation}
\bar B \gamma_\mu \gamma_5 B' \partial^\mu M \approx -i (m_B + m_{B'})\bar B \gamma_5 B' M
.
\end{equation}
The meson-baryon vertices relevant in the calculation of the $K$ meson exchange process are then given by
\begin{eqnarray}
{\cal L}_{mBB} 
&\approx & 
g_{K\Lambda N} 
\Bigl[
\bar p i\gamma_5 \Lambda K^+ 
+\bar n i\gamma_5 \Lambda K^0 
\Bigr]
\nonumber\\
&&
+g_{K\Sigma N} 
\Biggl[
\bar \Sigma^+ i\gamma_5 p \bar K^0 
+\frac{1}{\sqrt{2}} 
\Bigl(
\bar \Sigma^0 i\gamma_5 p K^-  
-\bar \Sigma^0 i\gamma_5 n \bar K^0
\Bigr)
\Biggr]
+ ({\rm h.c.})
\, ,
\end{eqnarray}
where $g_{K\Lambda N}=\frac{m_N+m_\Lambda}{2\sqrt{3} f_\pi } (D+3F) \approx 13.6$ and
$g_{K\Sigma N } = \frac{m_N +m_\Sigma}{\sqrt{2} f_\pi } (F-D) \approx -6.0$, with $D=0.81$ and $F=0.44$ (see Appendix \ref{sec:chiral_lagrangian}).
Here we have neglected the isospin splitting.

We now construct the CP-odd $K$ meson exchange nuclear force by considering the amplitude of the CP-odd $N-N$ scattering:
\begin{eqnarray}
i{\cal M}_{N-N}
&=&
\frac{i}{q^2 - m_K^2}
\Biggl[
\xi_{pp} \bar p i \gamma_5 p \cdot \bar pp
+\xi_{pn} \bar p i \gamma_5 p \cdot \bar nn 
+\xi_{np} \bar n i \gamma_5 n \cdot \bar pp
+\xi_{nn} \bar n i \gamma_5 n \cdot \bar nn 
\nonumber\\
&& \hspace{5em}
+ ( \xi_{\rm ch} \bar n i \gamma_5 p \cdot \bar pn  + {\rm h.c.})
\Biggr]
,
\end{eqnarray}
where $q$ is the momentum exchanged between two nucleons.
Note that the hermitian conjugate is only taken for the term with $\xi_{\rm ch}$.
The CP-odd $N-N$ couplings are given by
\begin{eqnarray}
\xi_{pp}
&=&
\frac{2 \, {\rm Re} (  a_{ p \Sigma^+} \, \bar g_{\bar K^0 p p} ) \, g_{K\Sigma N}}{m_N -m_\Sigma}
\nonumber\\
&=&
-\frac{2 J G_{yz} g_{K\Sigma N} \langle 0 | \bar q q + \bar ss | 0 \rangle
\langle p | \bar d d | p \rangle
\langle p | Q_2^{NR} | \Sigma^+ \rangle
 }{\sqrt{2} f_K (m_N -m_\Sigma )} 
,
\label{eq:xipp}
\\
\xi_{pn}
&=&
\frac{2 \, {\rm Re} (  a_{p \Sigma^+} \, \bar g_{\bar K^0 nn} ) \, g_{K\Sigma N}}{m_N -m_\Sigma}
\nonumber\\
&=&
-\frac{2 J G_{yz} g_{K\Sigma N} \langle 0 | \bar q q + \bar ss | 0 \rangle
\langle n | \bar d d | n \rangle
\langle p | Q_2^{NR} | \Sigma^+ \rangle
 }{\sqrt{2} f_K (m_N -m_\Sigma )} 
,
\label{eq:xipn}
\\
\xi_{np}
&=&
-\frac{\sqrt{2} \, {\rm Re} ( a_{n \Sigma^0} \, \bar g_{\bar K^0 pp} ) \, g_{K\Sigma N}}{m_N -m_\Sigma}
+\frac{2 \, {\rm Re} ( a_{n \Lambda} \, \bar g_{\bar K^0 pp} ) \, g_{K\Lambda N}}{m_N -m_\Lambda}
\nonumber\\
&=&
-\frac{2 J G_{yz} \langle 0 | \bar q q + \bar ss | 0 \rangle
\langle p | \bar d d | p \rangle}{\sqrt{2} f_K }
\Biggl[
- \frac{g_{K\Sigma N} \langle n | Q_2^{NR} | \Sigma^0 \rangle
 }{\sqrt{2} (m_N -m_\Sigma) } 
+
\frac{g_{K\Lambda N} \langle n | Q_2^{NR} | \Lambda \rangle
 }{m_N -m_\Lambda } 
\Biggr]
,
\ \ \ \ \ 
\label{eq:xinp}
\end{eqnarray}
\begin{eqnarray}
\xi_{nn}
&=&
-\frac{\sqrt{2} \, {\rm Re} ( a_{n \Sigma^0} \, \bar g_{\bar K^0 nn} ) \, g_{K\Sigma N}}{m_N -m_\Sigma}
+\frac{2 \, {\rm Re} (  a_{n \Lambda} \, \bar g_{\bar K^0 nn} ) \, g_{K\Lambda N}}{m_N -m_\Lambda}
\nonumber\\
&=&
-\frac{2 J G_{yz} \langle 0 | \bar q q + \bar ss | 0 \rangle
\langle n | \bar d d | n \rangle}{\sqrt{2} f_K }
\Biggl[
- \frac{g_{K\Sigma N} \langle n | Q_2^{NR} | \Sigma^0 \rangle
 }{\sqrt{2} (m_N -m_\Sigma) } 
+
\frac{g_{K\Lambda N} \langle n | Q_2^{NR} | \Lambda \rangle
 }{m_N -m_\Lambda } 
\Biggr]
,\ \ \ \ \ \ \ \ \ 
\label{eq:xinn}
\\
{\rm Re} ( \xi_{\rm ch} )
&=&
\frac{{\rm Re} ( a_{n\Sigma^0} \, \bar g_{K^- pn} ) \, g_{K\Sigma N}}{\sqrt{2} (m_N -m_\Sigma )}
+\frac{{\rm Re} ( a_{n \Lambda} \, \bar g_{K^- pn} ) \, g_{K\Lambda N}}{m_N -m_\Lambda}
\nonumber\\
&=&
-\frac{ J G_{yz} \langle 0 | \bar q q + \bar ss | 0 \rangle
\langle p | \bar u d | n \rangle}{\sqrt{2} f_K }
\Biggl[
\frac{g_{K\Sigma N} \langle n | Q_2^{NR} | \Sigma^0 \rangle
 }{\sqrt{2} (m_N -m_\Sigma) } 
+
\frac{g_{K\Lambda N} \langle n | Q_2^{NR} | \Lambda \rangle
 }{m_N -m_\Lambda } 
\Biggr]
.
\label{eq:xich}
\end{eqnarray}

In the isospin base, the general CP-odd one-kaon exchange nuclear force can be written as
\begin{eqnarray}
H_{P\hspace{-.5em}/\, T\hspace{-.5em}/\, }^K
& = &
 \frac{1}{2m_N} \bigg\{ 
\bar{G}_{K}^{(0')}\,{\mathbf \sigma}_{-}\cdot {\mathbf \nabla} {\cal Y}_K (r)
\nonumber \\
&& \hspace{3em}
+
\bar{G}_K^{(0)}\,{\mathbf \tau}_{1}\cdot {\mathbf \tau}_{2}\, {\mathbf \sigma}_{-}\cdot {\mathbf \nabla} {\cal Y}_K (r)
\nonumber \\
&& \hspace{3em} 
+\frac{1}{2} \bar{G}_K^{(1)}\,
( \tau_{+}^{z}\, {\mathbf \sigma}_{-} +\tau_{-}^{z}\,{\mathbf \sigma}_{+} )
\cdot {\mathbf \nabla} {\cal Y}_K (r)
\nonumber \\
&& \hspace{3em} 
+\frac{1}{2} \bar{G}_K^{(1')}\,
( -\tau_{+}^{z}\, {\mathbf \sigma}_{-} 
+\tau_{-}^{z}\,{\mathbf \sigma}_{+} )
\cdot {\mathbf \nabla} {\cal Y}_K (r)
\nonumber \\
&& \hspace{3em}
+\bar{G}_K^{(2)}\, (3\tau_{1}^{z}\tau_{2}^{z}- {\mathbf \tau}_{1}\cdot {\mathbf \tau}_{2})\,{\mathbf \sigma}_{-}\cdot {\mathbf \nabla}
{\cal Y}_K (r)
\ \bigg\}
,
\label{eq:KexchangeCPVhamiltonian}
\end{eqnarray}
where the notation is similar as that used in Eq. (\ref{eq:CPVhamiltonian}).
The kaon mass is $m_K = 495$ MeV.

Again, the CP-odd kaon exchange couplings can be converted from the charge base to the isospin base and vice versa.
If we define the transformation from the charge base to the isospin base as
\begin{equation}
\left(
\begin{array}{c}
\bar G^{(0')}_K \cr
\bar G^{(0)}_K \cr
\bar G^{(1)}_K \cr
\bar G^{(1')}_K \cr
\bar G^{(2)}_K \cr
\end{array}
\right)
=
A_K^{-1}
\left(
\begin{array}{c}
\xi_{pp} \cr
\xi_{pn} \cr
\xi_{np} \cr
\xi_{nn} \cr
\xi_{\rm ch} \cr
\end{array}
\right)
,
\end{equation}
the matrix $A_K$ will be given by
\begin{equation}
A_K =
\left(
\begin{array}{rrrrr}
\ 1 &1 & 1  & 1 &  2 \cr
1 &-1 & 1 & -1 &  -2 \cr
1 &-1 & -1 & 1 &  -2 \cr
1 & 1 & -1 & -1 &  2 \cr
0 & 2 & 0 & 0 &  -2  \cr
\end{array}
\right)
.
\end{equation}
We can then convert the CP-odd couplings of Eqs. (\ref{eq:xipp}), (\ref{eq:xipn}), (\ref{eq:xinp}), (\ref{eq:xinn}), and (\ref{eq:xich}) in the isospin base.
The explicit values of the coupling constants are
\begin{eqnarray}
\bar G^{(0')}_K 
&=& 
5.4 \times 10^{-16}
,
\label{eq:g0'k}
\\
\bar G^{(0)}_K 
&=& 
9.9 \times 10^{-18}
,
\label{eq:g0k}
\\
\bar G^{(1)}_K
&=& 
-1.3 \times 10^{-16}
,
\label{eq:g1k}
\\
\bar G^{(1')}_K
&=& 
-4.0 \times 10^{-17}
,
\label{eq:g1'k}
\\
\bar G^{(2)}_K
&=& 
-4.1 \times 10^{-21}
.
\label{eq:g2k}
\end{eqnarray}
We see that the couplings $\bar G^{(0')}_K$ and $\bar G^{(1)}_K$ are larger than the others.
This is because they receive contribution from the isoscalar nucleon scalar density [see Eq. (\ref{eq:nucleonscalardensity})], which is larger than other scalar baryon matrix elements.
The isotensor coupling $\bar G^{(2)}_K$ is suppressed since the hyperon-nucleon transition and the $|\Delta S| =1$ meson-baryon interaction we have used are both $|\Delta I| = \frac{1}{2}$.

\subsection{Electric dipole moments of the deuteron, $^3$H and $^3$He nuclei from the CP-odd nuclear force}
\label{sec:nuclearedm}

Now that we have the CP-odd nuclear force, we can now evaluate the nuclear EDM.
The nuclear EDM due to the nuclear polarization is defined as
\begin{eqnarray}
d_{A}
&=&
\sum_{i=1}^{A} \frac{e}{2} 
\langle \, \tilde A \, |\, (1+\tau_i^z ) \, {\cal R}_{iz} \, | \, \tilde A \, \rangle
,
\label{eq:nuclearedmpolarization}
\end{eqnarray}
where $|\, \tilde A\, \rangle$ is the polarized nuclear wave function, and $\tau^z_i$ is the isospin Pauli matrix.
${\cal R}_{iz}$ is the $z$-component of the coordinate of the nucleon in the center of mass frame of the nucleus.
The nuclear wave function must be a mixing of opposite parity states to yield a nonzero nuclear EDM.

Of course we should be aware that the nuclear EDM also receives contributions from the nucleon EDM.
This is given by
\begin{eqnarray}
d_A^{\rm (Nedm)} 
&=&
\sum_{i=1}^A
\frac{1}{2} \langle \, \tilde A\, |\, (d_p + d_n) \sigma_{iz} + (d_p - d_n) \tau_i^z \sigma_{iz} \, |\,  \tilde A\, \rangle
,
\end{eqnarray}
where $\sigma_{iz}$ is the spin operator, $d_p$ and $d_n$ the EDM of the proton and neutron, respectively.
The nucleon EDM contribution from the SM \cite{seng,mannel} may be comparable to that generated by the CP-odd nuclear force (\ref{eq:nuclearedmpolarization}), but here we are interested in the latter, since the effect of the single nucleon EDM is not enhanced by the nuclear effect, as the nucleus is not a relativistic system \cite{khriplovichbook,ginges,sandars1,sandars2,flambaumenhancement,yamanakanuclearedm}.
Rather, the nuclear spin matrix element cannot become larger than one for most of the stable nuclei, due to the strong pairing correlation.
It has also recently been pointed that Schiff's screening, although being incomplete, may occur in nuclear systems \cite{inoue}. 
The relative size between the nucleon EDM contribution and the polarization due to the CP-odd nuclear force will be discussed in the analysis.

As we have seen, the $|\Delta S| =1$ hadron level interactions of the SM generate pion-, eta- and kaon-exchange CP-odd nuclear forces.
Owing to their small couplings, the meson-exchange effect to the nuclear EDM is just given by the linear terms
\begin{equation}
d_A^{\rm (pol)}
=
\sum_{i} a_{A,\pi}^{(i)} \bar G_\pi^{(i)} 
+\sum_{j} a_{A,\eta}^{(j)} \bar G_\eta^{(j)} 
+\sum_{k} a_{A,K}^{(k)} \bar G_K^{(k)}
,
\end{equation}
where the linear coefficients $a_\pi^{(i)}$ $(i=0,1,2)$, $a_\eta^{(j)}$ $(j=0',1')$ and $a_K^{(k)}$ $(k=0',0,1,1',2)$ depend only on the nuclear structure.
The linear coefficients for the pion and eta exchange CP-odd nuclear forces are known from previous studies, and are shown in Table \ref{table:nuclearedm}.

\begin{table*}
\caption{
The EDM coefficients of the pion and eta exchange CP-odd nuclear forces \cite{yamanakanuclearedm}.
The linear coefficients of the CP-odd $N-N$ coupling $a_X$ ($X=\pi , \eta $) are expressed in unit of $10^{-2} e$ fm.
The sign $-$ denotes that the result vanishes in our setup.
}
\begin{center}
\begin{tabular}{l|ccc|cc|}
  &$a_\pi^{(0)}$ & $a_\pi^{(1)}$ & $a_\pi^{(2)}$& $a_\eta^{(0')}$ &$a_\eta^{(1')}$ \\ 
\hline
$^{2}$H & $-$ & $1.45 $ & $-$ &  $-$ & $0.157$ \\
$^{3}$He & $0.59$ & 1.08 & 1.68 &  $-5.77 \times 10^{-2}$ & 0.106 \\
$^{3}$H & $-0.59$ & 1.08 & $-1.70$ &  $5.80 \times 10^{-2}$ & 0.106 \\
$^{6}$Li & $-$ & 2.8 & $-$ & $-$ & 0.16 \\
$^{9}$Be & $-$ & $1.7$ & $-$ & $-$ & $- $ \\
\hline
\end{tabular}
\end{center}
\label{table:nuclearedm}
\end{table*}


The kaon-exchange CP-odd nuclear force has however not been studied in the ab initio approach so far.
Here we propose to calculate the nuclear EDM of the deuteron, $^3$H and $^3$He nuclei in the presence of the kaon exchange CP-odd nuclear force (\ref{eq:KexchangeCPVhamiltonian}).
As the ab initio method, we use the Gaussian expansion method \cite{hiyama}.
The Gaussian expansion method was successfully applied to the calculations of the EDM of the deuteron, the $^3$H, and $^3$He nuclei in a previous work \cite{yamanakanuclearedm}, and we expect that it also works well in this case.

The principle of the Gaussian expansion method is to solve the nonrelativistic Schr\"{o}dinger equation 
\begin{eqnarray}
( H - E ) \, \Psi_{JM,TT_z}  = 0 ,
\label{eq:schr7}
\end{eqnarray}
by diagonalizing the hamiltonian $H$ expressed in the gaussian basis.
The hamiltonian in this work is given by
\begin{equation}
H=\sum_a T_a+ \sum_{a,b} V_{{\rm A}v18 , ab}
+H_{P\hspace{-.5em}/\, T\hspace{-.5em}/\, , ab}^K
,
\label{eq:hamil7}
\end{equation}
with the kinetic energy operator $T$, the Argonne $v18$ nuclear force \cite{av18} $V_{{\rm A}v18 , ab}$, and the CP-odd kaon exchange nuclear force $H_{P\hspace{-.5em}/\, T\hspace{-.5em}/\, , ab}^K$ [see Eq. (\ref{eq:KexchangeCPVhamiltonian})], where explicit indices indicating the labels of the constituent nucleons $a$ and $b$ were written.
In this calculation, we do not include the effect of three-body forces \cite{3bodyforce1,3bodyforce2,3bodyforce3,3bodyforce4}.

For the deuteron, only one coordinate is needed.
We express the wave function of the relative coordinate as a superposition of the gaussian base
\begin{eqnarray}
\phi_{nlm}({\bf r})
&=&
N_{nl } r^l \, e^{-(r/r_n)^2}
Y_{lm}({\widehat {\bf r}})  \;  ,
\end{eqnarray}
with $N_{nl} $ the normalization constant, and the gaussian range parameters in the geometric progression
\begin{eqnarray}
      r_n
      &=&
      r_1 a^{n-1} \qquad \enspace
      (n=1 - n_{\rm max}) \; .
\end{eqnarray}
After diagonalizing the hamiltonian (\ref{eq:hamil7}) expressed in the gaussian basis, we take the ground state wave function and take the expectation value of the dipole operator as Eq. (\ref{eq:nuclearedmpolarization}) to obtain the deuteron EDM.
In our setup, the binding energies of the deuteron 2.22 MeV was correctly reproduced.
By factorizing the CP-odd nuclear coupling $\bar G_K^{(i)}$'s, we obtain the coefficients displayed in Table \ref{table:nuclearedmkaon}.

\begin{figure}[htb]
\begin{center}
\includegraphics[width=14cm]{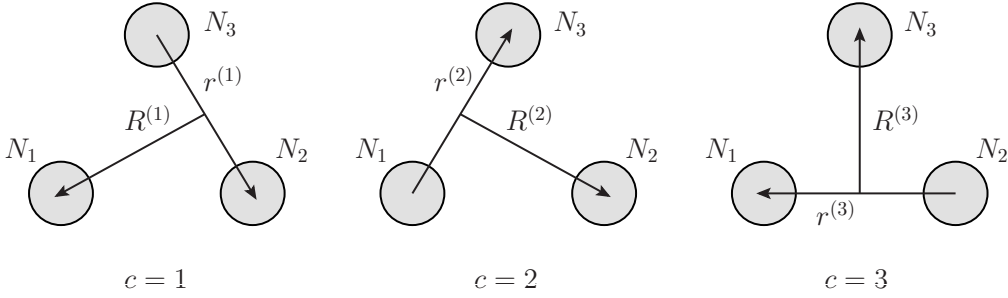}
\caption{\label{fig:jacobi}
The Jacobi coordinate for the 3-nucleon systems.
}
\end{center}
\end{figure}

For the three-nucleon systems, we need two coordinates to express the wave function.
Here we adopt three kinds of Jacobian coordinates shown in Fig. \ref{fig:jacobi} \cite{hiyama,3nucleon}.
The total wave function of the 3-nucleon systems is given as 
\begin{eqnarray}
\Psi_{JM, TT_z}(^3{\rm H}, ^3{\rm He})
&=&
 \sum_{c=1}^{3} \:
\sum_{nl, NL}
\sum_{T} \sum_{\Sigma} \sum_{S}
C^{(c)}_{nl,NL, \Sigma S, T}\: {\cal A} \Biggl[
\Bigl[  \eta^{(c)} ( T_c ) \otimes \eta'^{(c)} ({\scriptstyle \frac{1}{2}} ) \Bigr]_{I={\scriptstyle \frac{1}{2}}}
\nonumber  \\
&&  \times 
\Bigl[
[ \phi^{(c)}_{nl}({\bf r}_c) \otimes \psi^{(c)}_{NL}({\bf R}_c)]_\Lambda \otimes \, \bigl[ \chi^{(c)} ( S_c ) \otimes \chi'^{(c)} ({\scriptstyle \frac{1}{2}} ) \bigr]_\Sigma \Bigr]_{J={\scriptstyle \frac{1}{2}}, M}
\Biggr]
. \ \ \ \ \ \ 
\label{eq:he7lwf}
\end{eqnarray}
Here the operator $\cal{A}$ denotes the antisymmetrization of the whole system, and the arguments inside the large bracket are entangled so as to fulfill the antisymmetry in the interchange of each nucleon.
The spin function is given by $\chi$ and the isospin function by $\eta$.

The radial components of the wave function in the Gaussian expansion method for the two Jacobi coordinates are taken as \cite{3nucleon,hiyama,Hiyama2012ptep}
\begin{eqnarray}
\phi_{nlm}({\bf r})
&=&
r^l \, e^{-(r/r_n)^2}
Y_{lm}({\widehat {\bf r}})  \;  ,
\nonumber \\
\psi_{NLM}({\bf R})
&=&
R^L \, e^{-(R/R_N)^2}
Y_{LM}({\widehat {\bf R}})  \;  ,
\end{eqnarray}
where the gaussian range parameters obey the following geometric progression:
\begin{eqnarray}
      r_n
      &=&
      r_1 a^{n-1} \qquad \enspace
      (n=1 - n_{\rm max}) \; ,
\nonumber\\
      R_N
      &=&
      R_1 A^{N-1} \quad
     (N \! =1 - N_{\rm max}).
\end{eqnarray}
The results of our work have well converged for the angular momentum space $l, L, \Lambda \leq 2$.
In our setup of the realistic nuclear force, we have found 6.93 MeV and 7.63 MeV for the binding energies of the $^3$He and $^3$He nuclei, respectively.
Again by factorizing the CP-odd nuclear coupling $\bar G_K^{(i)}$'s, we obtain the coefficients in Table \ref{table:nuclearedmkaon}.

By comparing Table \ref{table:nuclearedm} and Table \ref{table:nuclearedmkaon}, we see that the coefficients of the kaon exchange CP-odd nuclear force are smaller than those of the pion exchange.
This result is quite natural since the kaon is heavier than the pion.
See also the similarity between the kaon exchange coefficients $\bar G_K^{(0')}$ and $\bar G_K^{(1')}$ and the eta exchange coefficients $\bar G_\eta^{(0')}$, $\bar G_\eta^{(1')}$of Table \ref{table:nuclearedm}, which is convincing since the kaon and the $\eta$ meson have close masses.
Before the coefficients of Table \ref{table:nuclearedmkaon} were available, the kaon exchange CP-odd potential was approximated as a contact interaction \cite{smCPVNN}, with the factor $\frac{1}{m_K^2}$ instead of the Yukawa function.
If we also assume the contact interaction for the pion exchange potential, the ratio between the linear coefficients of the pion and kaon exchange CP-odd nuclear couplings is $ \frac{a_\pi^{(i)}}{a_K^{(i)}} = \frac{m_K^2 }{ m_\pi^2} \sim 12$ ($i=0,1,2$). 
On the other hand, the result of our ab initio evaluation shows a typical ratio $\frac{a_\pi^{(i)}}{a_K^{(i)}} \sim 7$ (compare values of Tables \ref{table:nuclearedm} and \ref{table:nuclearedmkaon}).
We see that the kaon exchange contribution evaluated in the Gaussian expansion method is larger than that given by the simple contact interaction estimate.
This shows that the kaon exchange process has enough long range contribution so that it cannot be described by the short range contact interaction approximation.
This difference is not negligible if both pion and kaon exchange contributions are considered simultaneously, and shows the importance of the ab initio evaluation.

\begin{table*}
\caption{
The EDM coefficients for $K$ meson exchange process.
The linear coefficients $a_K$ are expressed in unit of $10^{-2} e$ fm.
The sign $-$ denotes that the result vanishes in our setup.
}
\begin{center}
\begin{tabular}{l|ccccc|}
  &$a_K^{(0')}$ & $a_K^{(0 )}$ &$a_K^{(1)}$ &$a_K^{(1')}$ &$a_K^{(2)}$\\ 
\hline
$^{2}$H &  $-$ & $-$ & 0.202 & $0.202$ & $-$  \\
$^{3}$He & $-7.42 \times 10^{-2}$& $8.72 \times 10^{-2}$ & $0.173$ & 0.135 & $0.249 $ \\
$^{3}$H & $7.45 \times 10^{-2}$ & $-8.81 \times 10^{-2}$ & $0.175$ & 0.135 & $-0.254$ \\
\hline
\end{tabular}
\end{center}
\label{table:nuclearedmkaon}
\end{table*}

We should add a comment on the coefficients of $^6$Li and $^9$Be EDMs which have not been evaluated for the kaon exchange nuclear force.
In Ref. \cite{yamanakanuclearedm}, the EDMs of those nuclei were evaluated in the cluster model which uses the folding of the CP-odd nuclear force with the nuclear density of the $\alpha$ cluster.
There it was pointed that the result for the CP-odd $N-N$ interaction with the exchange of mesons heavier than the pion may be an overestimation, due to the averaging of the short range contribution by the folding. 
This is the reason why we have not calculated the kaon exchange coefficients for $^6$Li and $^9$Be.


\section{Results and analysis}

From our calculation, we obtain the following nuclear EDM results
\begin{eqnarray}
d_{^2{\rm H}}^{\rm (pol)}
&=&
2.2 \times 10^{-31}e\, {\rm cm}
,
\label{eq:deuteronedm}
\\
d_{^3{\rm He}}^{\rm (pol)}
&=&
2.2 \times 10^{-31}e\, {\rm cm}
,
\label{eq:3heedm}
\\
d_{^3{\rm H}}^{\rm (pol)}
&=&
1.0 \times 10^{-31}e\, {\rm cm}
,
\label{eq:3hedm}
\end{eqnarray}
which are the sum of the pion exchange contribution 
\begin{eqnarray}
d_{^2{\rm H}}^\pi
&=&
2.6 \times 10^{-31}e\, {\rm cm}
,
\\
d_{^3{\rm He}}^\pi
&=&
2.9 \times 10^{-31}e\, {\rm cm}
,
\\
d_{^3{\rm H}}^\pi
&=&
1.0 \times 10^{-31}e\, {\rm cm}
,
\end{eqnarray}
the eta exchange contribution 
\begin{eqnarray}
d_{^2{\rm H}}^\eta
&=&
-5.7 \times 10^{-33}e\, {\rm cm}
,
\\
d_{^3{\rm He}}^\eta
&=&
-1.5 \times 10^{-33}e\, {\rm cm}
,
\\
d_{^3{\rm H}}^\eta
&=&
-6.2 \times 10^{-33}e\, {\rm cm}
,
\end{eqnarray}
and the $K$ meson exchange contribution
\begin{eqnarray}
d_{^2{\rm H}}^K
&=&
-3.5
\times 10^{-32}e\, {\rm cm}
,
\\
d_{^3{\rm He}}^K
&=&
-6.8
\times 10^{-32}e\, {\rm cm}
,
\\
d_{^3{\rm H}}^K
&=&
1.0
\times 10^{-32}e\, {\rm cm}
.
\end{eqnarray}
We see that the pion exchange contribution is the leading one.
The kaon exchange effect is also sizable, and acts destructively for the deuteron and $^3$He EDMs.
The $\eta$ meson exchange contribution was found to be subleading.
Let us try to understand the order of magnitude of our result.
As we have seen in the introduction, the naive estimate of the EDM of light nuclei using dimensional analysis gives $d_A \sim O( \frac{\alpha_s}{4\pi} G_F^2 J \Lambda_{\rm QCD}^3 ) \sim 10^{-32}e$ cm, which is one order of magnitude below our values.
The enhancement has occurred in the renormalization group evolution of the Wilson coefficient of the penguin operator $Q_6$.
The renormalization group evolution of the $|\Delta S| = 1$ operator from the electroweak scale to the hadronic scale enhances the Wilson coefficient of $Q_6$ by a factor of 40 [compare Eqs. (\ref{eq:wilsoncoefinit}) and (\ref{eq:zytilde})], which is much larger than the growth of the strong coupling constant $\frac{\alpha_s (\mu = 1 \, {\rm GeV}) }{\alpha_s (\mu = m_w)} \sim 5$.
Our result shows that including the effect of renormalization group evolution is indispensable in the analysis of hadronic EDM.

Let us now compare the effects of exchanged mesons.
In this work, we have found that the pion exchange contribution is dominant.
The eta meson exchange contribution is smaller than the pion exchange by about one order of magnitude, although they have CP-odd nuclear couplings of similar size [compare Eqs. (\ref{eq:g0eta}) and (\ref{eq:g1eta}) with (\ref{eq:g0pi}), (\ref{eq:g1pi}), and (\ref{eq:g2pi})].
This can be explained by the small linear coefficients relating the CP-odd nuclear couplings to the nuclear EDM (see Table \ref{table:nuclearedm}).
The kaon exchange contribution, however, although not being leading, is competitive.
The kaon exchange CP-odd nuclear force has smaller linear coefficients than the pion exchange process, by roughly a factor of 7.
Its competitivity can be explained by the large couplings $\bar G_K^{(0')}$ (\ref{eq:g0'k}) and $\bar G_K^{(1)}$ (\ref{eq:g1k}), which receive contribution from the isoscalar nucleon scalar density $\langle N | \bar uu + \bar dd | N \rangle \sim 10$ [see Eq. (\ref{eq:nucleonscalardensity})].
The other couplings, including those for pion and eta exchanges, issue from other scalar densities which are between one and two.
Although this sensitive dependence on the scalar density is due to the factorization model, our result shows that the kaon exchange CP-odd nuclear force must be carefully analyzed in each hadronic system.

We also show the value of the EDM of heavier light nuclei by only assuming the pion exchange contribution.
The EDM of the $^6$Li and $^9$Be nuclei can be estimated as
\begin{eqnarray}
d_{^6{\rm Li}}^\pi
&=&
5.0 \times 10^{-31}e\, {\rm cm}
,
\\
d_{^9{\rm Be}}^\pi
&=&
3.1 \times 10^{-31}e\, {\rm cm}
.
\end{eqnarray}
We have used the result of Ref. \cite{yamanakanuclearedm} (see Table \ref{table:nuclearedm}), where the EDM of those nuclei was evaluated in the cluster model \cite{clusterreview1,clusterreview2,clusterreview3}.
There it was pointed that the $^6$Li has a large sensitivity on the isovector CP-odd nuclear force, due to its cluster structure.
This enhancement also enlarges the SM contribution, as we can see from the result.
As we have mentioned in the previous section, the EDM due to the kaon exchange CP-odd nuclear force cannot be discussed, because the folding used in Ref. \cite{yamanakanuclearedm} will destroy the information of the short range interaction.
The study of the kaon exchange contribution to $^6$Li and $^9$Be is left for the ab initio analysis.
The EDM of light nuclei generated in the SM shows values around $10^{-31}e$ cm.
These are beyond the prospective experimental sensitivity [$O(10^{-29})e$ cm] of the planned measurement using storage rings \cite{storage5,bnl}, but not far from it.
As the many-body effect may enhance the nuclear EDM for heavier nuclei, it is important to carefully inspect the SM contribution in nuclear systems.

The enhancement of the CP violation due to the many-body effect may also be relevant in nuclear reactions as well as nuclear decays \cite{gudkovreview}.
It has been suggested in Ref. \cite{bunakov} that the scattering of polarized neutron through polarized nuclear target may enhance the effect of CP-odd nuclear force by a factor of $10^6$, and the experimental realization of its measurement has extensively been studied \cite{kabir1,stodolsky,kabir2,kabir3,lamoreaux,skoi,lukashevich,bowman}.
Recently, the measurement of T-violating neutron scattering using the spallation neutron source is being prepared at Oak Ridge National Laboratory Laboratory and at J-PARC.
Quantitative theoretical studies of the  CP violating neutron scattering off nuclear target have also begun, starting from the simplest neutron-deuteron system \cite{song2,song3,uzikov}.
Here the SM contribution will also be an important systematics which have to be controlled.
When the nuclear level formulae of CP violating observables for heavy nuclear targets will be available, the SM effect will be given by combining them with the CP-odd nuclear couplings of Eqs. (\ref{eq:g0pi}), (\ref{eq:g1pi}), (\ref{eq:g2pi}), (\ref{eq:g0eta}), (\ref{eq:g1eta}), (\ref{eq:g0'k}), (\ref{eq:g0k}), (\ref{eq:g1k}), (\ref{eq:g1'k}), and (\ref{eq:g2k}).

As another quantities of interest, we have the SM contribution to the EDM of $^{129}$Xe, $^{199}$Hg, $^{211}$Rn, and $^{225}$Ra atoms generated through the nuclear Schiff moment.
The EDM of a nucleus inside neutral atoms is completely screened if it is nonrelativistic and point-like \cite{schiff}.
The nuclear finite volume effect can however polarize the atomic system through the nuclear Schiff moment, defined as \cite{ginges}
\begin{eqnarray}
S_A 
&\equiv &
\frac{e}{30} \sum_{p=1}^{Z}
\langle \, \tilde A \, |\, [ 3 {\cal R}_{p}^2 - 5 \langle r_{\rm ch}^2 \rangle ] {\cal R}_{pz} \, | \, \tilde A \, \rangle
\nonumber\\
&=&
\sum_{i} a_{A,\pi}^{S(i)} \bar G_\pi^{(i)} 
,
\end{eqnarray}
where $\langle r_{\rm ch}^2 \rangle$ is the mean squared charge radius of the nucleus $A$.
Here the linear coefficients $a_\pi^{S(i)}$ $(i=0,1,2)$ depend only on the nuclear structure, and their values are shown in Table \ref{table:xeschiffmoment} for $^{129}$Xe, and in Table \ref{table:schiffmoment} for $^{199}$Hg, $^{211}$Rn, and $^{225}$Ra atoms.
The atomic EDM receives the following contribution from the nuclear Schiff moment
\begin{equation}
d^S_A 
=
K^S_A S_A
,
\end{equation}
where the $K^S_A$ is a coefficient derived from the atomic physics (see Table \ref{table:diamagnetic}).
We neglect the intrinsic nucleon EDM contribution for the same reason as that for the EDM of light nuclei.

Again by only assuming the pion exchange contribution, the atomic EDMs in the SM are given by
\begin{eqnarray}
d_{\rm Xe}^{S}
&=&
2.2 \times 10^{-36}e\, {\rm cm}
,
\\
d_{\rm Hg}^{S}
&=&
-4.2 \times 10^{-35}e\, {\rm cm}
,
\\
d_{\rm Rn}^{S}
&=&
2.0 \times 10^{-34}e\, {\rm cm}
,
\\
d_{\rm Ra}^{S}
&=&
-7.4 \times 10^{-32}e\, {\rm cm}
.
\end{eqnarray}
The atomic EDM generated by the nuclear Schiff moment is small in the SM, well below the current experimental sensitivity \cite{rosenberry,griffith,parker}.
We should note that the atomic EDM receives other contributions from the SM such as the EDM of the electron or the atomic polarization due to the P, CP-odd electron-nucleon interaction.
However, as their source of CP violation is not hadronic, we do not consider them in this work.

\begin{table}
\begin{center}
\caption{\label{table:xeschiffmoment}Linear coefficients of the dependence of the Schiff moment of $^{129}$Xe on CP-odd nuclear couplings calculated in the mean field approach \cite{dmitriev} and in the shell model \cite{yoshinaga1,yoshinaga2,yoshinaga3}.
The constants are in unit of $e$ fm$^3$.
Note that the sign convention of Ref. \cite{dmitriev} is different from that of Refs. \cite{yoshinaga1,yoshinaga2,yoshinaga3}.
In this work we use the latest result of Ref. \cite{yoshinaga3}
}

\vspace{0.5em}

\begin{tabular}{lccc}
\hline
$^{129}$Xe& $a^{S(0)}_{\rm Xe}$\ \  &$a^{S(1)}_{\rm Xe}$ \ \ &$a^{S(2)}_{\rm Xe}$\ 
\\  
\hline
Dmitriev \cite{dmitriev} & 0.008 & 0.006 & 0.009  \\
Shell model \cite{yoshinaga1} & $5.07 \times 10^{-4}$ & $3.99 \times 10^{-4}$ & $1.89 \times 10^{-3}$ \\
Shell model \cite{yoshinaga3} & $2.74 \times 10^{-3}$ & $7.9 \times 10^{-4}$ & $2.00 \times 10^{-3}$ \\
\hline
\end{tabular}
\end{center}

\end{table}

\begin{table}
\begin{center}
\caption{\label{table:schiffmoment}Linear coefficients $a^{S(i)}$ of the dependence of the nuclear Schiff moment on CP-odd nuclear couplings in unit of $e$ fm$^3$ for the $^{199}$Hg, $^{211}$Rn \cite{hfodd,ban}, and $^{225}$Ra \cite{hfodd,dobaczewski} atoms.
The labels HB and HFB denote that the calculations were made in the Hartree-Fock and Hartree-Fock-Bogoliubov approximations, respectively.
The last line of each table denotes the values we use in this work.
}

\vspace{0.5em}

\begin{tabular}{lcccc}
\hline
$^{199}$Hg \cite{ban} & $a^{(0)}_{\rm Hg}$\ \  &$a^{(1)}_{\rm Hg}$ \ \ &$a^{(2)}_{\rm Hg}$\ \ 
\\ 
\hline
SkM$^*$ (HFB) \cite{skm} & 0.041 & $-$0.027 & 0.069 
\\
SLy4 (HFB) \cite{sly4} & 0.013 & $-$0.006 & 0.024 
\\
SLy4 (HF) \cite{sly4} & 0.013 & $-$0.006 & 0.022 
\\
SV (HF) \cite{sv} & 0.009 & $-$0.0001 & 0.016 
\\
SIII (HF) \cite{sv} & 0.012 & 0.005 & 0.016 
\\
\hline
Average&0.018&$-$0.007 &0.029 
\\
\hline
\end{tabular}

\vspace{0.5em}

\begin{tabular}{lcccc}
\hline
$^{211}$Rn \cite{ban} & $a^{(0)}_{\rm Rn}$\ \  &$a^{(1)}_{\rm Rn}$ \ \ &$a^{(2)}_{\rm Rn}$\ \ 
\\ 
\hline
SkM$^*$ \cite{skm} & 0.042 & $-$0.028 & 0.078 
\\
SLy4 \cite{sly4} & 0.042 & $-$0.018 & 0.071 
\\
SIII \cite{sv} & 0.034 & $-$0.0004 & 0.064 
\\
\hline
Average&0.039&$-$0.0015 &0.071 
\\
\hline
\end{tabular}

\vspace{0.5em}

\begin{tabular}{lcccc}
\hline
$^{225}$Ra \cite{dobaczewski} & $a^{(0)}_{\rm Ra}$\ \  &$a^{(1)}_{\rm Ra}$ \ \ &$a^{(2)}_{\rm Ra}$\ \ 
\\ 
\hline
SkM$^*$ \cite{skm} & $-$ 4.7 & 21.5 & $-$11.0 
\\
SLy4 \cite{sly4} & $-$ 3.0 & 16.9 & $-$8.8 
\\
SIII \cite{sv} & $-$ 1.0 & 7.0 & $-$3.9 
\\
\hline
SkO' \cite{sko} & $-$ 1.5& 6.0 &$-$4.0 
\\
\hline
\end{tabular}

\end{center}

\end{table}

\begin{table*}
\begin{center}
\caption{\label{table:diamagnetic}
The linear coefficients for $^{129}$Xe, $^{199}$Hg, $^{211}$Rn, and $^{225}$Ra atoms relating the nuclear Schiff moment and the EDM of atoms, in unit of cm/fm$^3$.
}
\vspace{0.5em}
\begin{tabular}{lc}
\hline
  &$K^S_a $ (cm/fm$^3$) \\ 
\hline
$^{129}$Xe \cite{flambaumdiamagnetic,singh1,Ramachandran} & $3.8 \times 10^{-18} $  \\
$^{199}$Hg \cite{flambaumdiamagnetic,lathahg,radziute,singh2} & $- 2.6 \times 10^{-17} $ \\
$^{211}$Rn \cite{flambaumdiamagnetic,sahoo} & $3.3 \times 10^{-17} $ \\
$^{225}$Ra \cite{flambaumdiamagnetic,Ramachandran,radziute,singh3} & $- 8.8 \times 10^{-17} $ \\
\hline
\end{tabular}
\end{center}

\end{table*}

Now we have to investigate the theoretical uncertainties of this work.
In our calculation the main sources of uncertainties are:
\begin{itemize}
\item
The electroweak contribution to the renormalization of the $|\Delta S| =1$ four-quark interactions.

\item
The renormalization of the $|\Delta S| =1$ four-quark interactions near the hadronic scale.

\item
The CP-odd $m\bar B B'$ hadron matrix elements $\langle m B | {\cal H}_{ eff} | B' \rangle$.

\item
The momentum dependence of the CP-odd $m\bar B B'$ vertex.

\item
Weak hyperon-nucleon transition.

\item
Uncertainty due to the nuclear force.

\item
Systematics due to other nucleon level contributions (single nucleon EDM, CP-odd meson exchange current, etc).

\item
The nuclear structure involving hyperons in the intermediate state.

\end{itemize}

The first important theoretical uncertainty is that due to the renormalization.
In this work, we have not considered the renormalization group evolution of the electroweak sector of the $|\Delta S| =1$ four-quark interaction \cite{burasreview,anatomyb}, but its effect is smaller than that of the strong interaction by more than one order.
The most important uncertainties in the renormalization group evolution are the uncertainty in the choice of the hadronic scale and the nonperturbative effect.
To visualize the dependence on the choice of the hadronic scale, we have calculated the EDM of light nuclei for $\mu = 2$ GeV and $\mu = 0.6$ GeV.
For $\mu = 2$ GeV, we have obtained
\begin{eqnarray}
d_{^2{\rm H}}^{\rm (pol)} 
&=&
2.6 \times 10^{-31}e\, {\rm cm}
,
\\
d_{^3{\rm He}}^{\rm (pol)}
&=&
2.6 \times 10^{-31}e\, {\rm cm}
,
\\
d_{^3{\rm H}}^{\rm (pol)}
&=&
1.2 \times 10^{-31}e\, {\rm cm}
,
\end{eqnarray}
and for $\mu = 0.6$ GeV
\begin{eqnarray}
d_{^2{\rm H}}^{\rm (pol)}
&=&
1.2 \times 10^{-31}e\, {\rm cm}
,
\\
d_{^3{\rm He}}^{\rm (pol)}
&=&
1.4 \times 10^{-31}e\, {\rm cm}
,
\\
d_{^3{\rm H}}^{\rm (pol)}
&=&
3.8 \times 10^{-32}e\, {\rm cm}
.
\end{eqnarray}
We see that the change of the scale from $\mu = 1 $ GeV to 2 GeV does not alter the order of the EDMs, whereas the result is smaller by about a factor of two than Eqs. (\ref{eq:deuteronedm}), (\ref{eq:3heedm}), and (\ref{eq:3hedm}) for $\mu =0.6$ GeV.
Of course the final observables must not in principle depend on the choice of this scale.
In our work, we are however calculating hadron matrix elements in the factorization model where the low energy factorization scale dependence cannot be controled.
Moreover, near this energy scale, the nonperturbative effect to the renormalization should become important.
The behavior of the Wilson coefficients at low energy scale can be accessed with lattice QCD, where nonperturbative renormalization is possible.
This calculation should be associated with the calculation of the $|\Delta S| =1$ hadron matrix elements.

The second important class of theoretical uncertainty is that due to the evaluation of the hadron matrix elements and hadron level effective interactions.
In this work, the calculation of the $|\Delta S| = 1$ effective interaction at the hadron level was done in the factorization model, for which the systematic error is unknown, certainly exceeding 100\%.
To calculate them quantitatively, the evaluation using lattice QCD is absolutely required.
There are currently no lattice QCD results available, but the calculations of the CP violating $K \rightarrow \pi \pi$ process which involves the same $|\Delta S| =1$ four-quark interactions at the elementary level are currently extensively done \cite{blum1,boucaud,giusti,blum2,blum3,bai}.
We can therefore expect similar techniques to be applied in the effective hadron level processes required for the study of the nuclear EDM in the SM.
The first target is the P-odd hadron matrix elements  $\langle m B | {\cal H}_{eff}^{P\hspace{-.5em}/\,} | B' \rangle$, which probably yields the most important amount of theoretical uncertainty in this work.
These matrix elements require the evaluation of 4-point functions in lattice QCD, as for the $K \rightarrow \pi \pi$ process.
The second target is the hyperon-nucleon interaction.
In our work, its evaluation has been done using the quark model.
Although the fit of the parameters successfully reproduce the hyperon decay \cite{hiyamahyperon-nucleon}, here again the lattice QCD simulation may provide a more accurate matrix element \cite{brambilla}.

We should also note that to control the systematics, we have to include the effect of excited strange baryons in the intermediate state.
The lowest excited state (with the same total angular momentum) of $\Lambda$ is $\Lambda$(1405), and that for $\Sigma$ is $\Sigma$(1660).
$\Lambda$(1405) is an opposite parity excited state ($1/2^-$), so the hyperon-nucleon transition must violate parity to obtain it.
This procedure will however require the $|\Delta S|=1$ CP-odd meson-baryon vertex to be parity conserving, and consequently be a vector meson exchange process, in the one-meson exchange factorization model.
The vector meson exchange CP-odd nuclear force contributes two orders of magnitude less than the pion exchange one, so it is subdominant.
$\Sigma$(1660) has the same parity as the nucleon ($1/2^+$), so it contributes in the same way as the octet baryons relevant in our work.
Although requiring an excitation of $m_{\Sigma (1660)} - m_N \simeq 700$ MeV, the simple kinematical suppression compared with the octet hyperon due to the mass shift is larger than 1/3, so its effect may potentially be important.

In this work, the CP-odd nuclear force was treated as a combination of the $|\Delta S| = 1$ meson-nucleon interactions, the CP-even $|\Delta S| = 0$ chiral meson-baryon interactions, and the hyperon-nucleon transition.
This CP-odd nuclear force is reducible to the $|\Delta S| = 1$ $NN \rightarrow YN$ ($Y$ is a hyperon) interaction and the hyperon-nucleon transition.
The $NN \rightarrow YN$ interactions were discussed in the context of the decay of hypernuclei \cite{inoue1,inoue2,sasaki1,sasaki2}.
If $NN \rightarrow YN$ interactions are simulated in lattice QCD, all informations of intermediate states such as the momentum dependence of the $|\Delta S| = 1$ and $|\Delta S| = 0$ meson-baryon interactions, off-shell effects, and contributions beyond one-meson exchange can be considered all at once.
Recently, it has become possible to extract the nuclear potentials from first principle in lattice QCD \cite{hal1,hal2}.
If this technique is applicable to the $NN \rightarrow YN$ potentials, we will experience much progress in the calculation of the nuclear CP violation in the SM.

The third class of theoretical uncertainty is that related to the inputs of the nuclear level calculation.
The first point is the quality of the CP-even realistic nuclear force.
In our work, we have used the Argonne $v18$ potential as a realistic nuclear force.
From the result of previous works \cite{liu,stetcu,song,bsaisou,bsaisou2}, the nuclear EDM agrees well for the deuteron, $^3$He, and $^3$H, at least within the level of 10\%.
The uncertainty due to the three-body force \cite{3bodyforce1,3bodyforce2,3bodyforce3,3bodyforce4} is also within the same level \cite{stetcu,song,bsaisou,bsaisou2}.

For the uncertainty related to the CP-odd nuclear level input, we have the intrinsic nucleon EDM contribution to the nuclear EDM.
As we have seen in the previous section, the nucleon EDM contribution to the nuclear EDM is at most one, since there are no multiple valence nucleons in the open shell \cite{yamanakanuclearedm}.
Moreover, nuclear systems are nonrelativistic, so we cannot expect any relativistic enhancement of the nucleon EDM effect.
Recently there have been two evaluations of the nucleon EDM.
The first one discusses the long-distance contribution to the nucleon EDM in the chiral approach \cite{seng}.
The predicted value is 
\begin{equation}
1 \times 10^{-32} e\, {\rm cm}
<
\{ |d_n| , |d_p| \}
<
6 \times 10^{-32} e \, {\rm cm}
,
\end{equation}
which is smaller than the CP-odd nuclear force contributions to the nuclear EDMs calculated in this work (\ref{eq:deuteronedm}), (\ref{eq:3heedm}), and (\ref{eq:3hedm}).
The second work has estimated the short distance contribution generated through loop-less effect \cite{mannel}.
The result is
\begin{equation}
d_n \sim 1.4 \times 10^{-31}e\, {\rm cm}
,
\label{nucleonedmshort}
\end{equation}
with typical parameters.
This value is comparable to our results (\ref{eq:deuteronedm}), (\ref{eq:3heedm}), and (\ref{eq:3hedm}).
It should be noted that Ref. \cite{mannel} has estimated the nucleon matrix element involving operators with three quark bilinears like $\bar u \gamma^\alpha (1-\gamma_5) s \cdot \bar s \gamma_\alpha \gamma^\mu \gamma_\beta (1-\gamma_5) d \cdot \bar d \gamma^\beta (1-\gamma_5) u$.
In the vacuum saturation approximation, only the isoscalar scalar density is enhanced.
Here we have however no contribution from it, so there is no reason for the nucleon matrix elements to be enhanced.
Rather, we can expect significant suppression from the gluon dressing effect for operators involving the axial and tensor currents \cite{yamanakasde1,yamanakasde2,pitschmann}.
Moreover, these operators involve strange quarks.
It is currently known that the nucleon matrix elements of strange quark charges are suppressed from lattice QCD analyses \cite{jlqcd2,jlqcd3,qcdsf1,young1,durr1,dinter,milc1,qcdsf2,engelhardt,gong,etm1,etm2,junnarkar,qcdsf3,bhattacharya2,bhattacharya3,jlqcd4}.
The estimated value (\ref{nucleonedmshort}) should be considered as an upper limit.

There is also another CP-odd input we have not considered, which is the contribution from the CP-odd meson exchange current.
In previous works, those have been estimated to be about 10\% of the total contribution \cite{liu,stetcu}.
The exchange current contribution to the magnetic moment is known to be large for some nuclei \cite{pastore1,pastore2,pastore3,pastore4}, so the inspection of the exchange current is necessary in the evaluation of the EDM of each nucleus.

The final systematics is the nuclear level long distance effect.
In our work, we do not have considered the dynamical effect of the hyperons appearing through $|\Delta S|=1$ processes.
In the real case, a $|\Delta S|=1$ process and its conjugate do not occur almost successively, since the $NN \rightarrow YN$ interaction and the hyperon-nucleon transition are very rare processes.
This means that the off-shell hyperon will sufficiently interact with other pieces of the nucleus, once created.
This effect is a dynamical long range effect that was not included in our work.
The evaluation of this contribution requires the calculation of every hypernuclei which may couple to the nucleus in question through the $|\Delta S| = 1$ interactions.
We have no idea of how important the correction is, as the EDM strongly depends on the structure of the ground and opposite parity excited states.
To obtain the true value of the SM contribution to the nuclear EDM, the mixing of the nucleus in question with hypernuclei through the $|\Delta S|=1$ process has to be evaluated.
This will be the subject of the future work.


\section{Summary}

In this paper, we have calculated the SM effect to the nuclear EDM of the deuteron, $^3$H and $^3$He in the factorization model by considering the  long distance effect of the $|\Delta S| =1$ processes as the leading contribution.
At the nuclear level, we have calculated the contribution of the light pseudoscalar meson ($\pi$, $\eta$ and $K$) exchange CP-odd nuclear force.
In particular, the effect of the kaon exchange to the nuclear EDM was evaluated for the first time, using the ab initio Gaussian expansion method.
As a result, we have obtained values which lie around $O(10^{-31})e$ cm, which is not far from the reach of the prospective sensitivity of the next generation experiment using storage rings [$O(10^{-29})e$ cm].
The predicted value is larger than the naive estimation $d_A \sim O(10^{-32})e$ cm, due to the enhancement of the Wilson coefficient of the $|\Delta S|=1$ four-quark operator $Q_6$ in the renormalization group evolution from $\mu = m_W$ to the hadronic scale $\mu =1 $ GeV.

We have also shown that the pion exchange CP-odd nuclear force gives the largest contribution to the nuclear EDM.
In the factorization approach, the kaon exchange contribution is however not small, due to the large value of the isoscalar scalar density of the nucleon.
Moreover, the relative size between the kaon exchange contribution to the EDMs of the deuteron, $^3$He, and $^3$H against that of the pion exchange, calculated in the Gaussian expansion method, shows linear coefficients larger than the contact interaction estimates.
This result indicates that the kaon exchange contribution is still important in the evaluation of the nuclear EDM.
The $\eta$ meson exchange was found to have smaller effect on the nuclear EDM than the pion exchange by about one order of magnitude, due to the absence of enhancement mechanism.

We also have estimated for the first time the nuclear EDM of heavier light nuclei, $^6$Li and $^9$Be, and the nuclear Schiff moment contribution to the EDM of $^{129}$Xe, $^{^199}$Hg, $^{211}$Rn, and $^{225}$Ra atoms in the SM, by only considering the pion exchange CP-odd nuclear force.
For $^6$Li and $^9$Be, the EDM is larger than those of the deuteron, $^3$He, and $^3$H, due to the cluster structure.
For the atomic EDMs, we have found values which are well below the current experimental sensitivity.

We have also investigated the theoretical uncertainty in this calculation.
The largest error comes mainly from the evaluation of the CP-odd hadron matrix elements.
We expect the lattice QCD simulation to determine them in the future, although this is a difficult challenge.
Moreover, the systematics due to the mixing of the nucleus with hypernuclei through $|\Delta S|=1$ is so far completely unknown.
The improvement of the theoretical accuracy is an important future subject in the study of the nuclear EDM in the SM.

\acknowledgments

The authors thank T. Hatsuda, T. Hyodo, S. Pastore, E. Teruya and N. Yoshinaga for useful discussions and comments.
The nuclear level calculation was performed using the supercomputer HITACHI SR16000 of the Yukawa Institute for Theoretical Physics.
This work is supported by the RIKEN iTHES Project.


\appendix

\section{Renormalization group equation of the Wilson coefficients of the $|\Delta S| = 1$ four-quark interactions \label{sec:penguinrge}}

In the next-to-leading logarithmic order, the running matrix of the Wilson coefficients of $|\Delta S| = 1$ four-quark operators [see Eqs. (\ref{eq:q1}), (\ref{eq:q2}), (\ref{eq:q3}), (\ref{eq:q4}), (\ref{eq:q5}), and (\ref{eq:q6})] is a $6\times 6$ matrix
\begin{equation}
\hat{U}_{N_f} (\Lambda_1 , \Lambda_2 )
=
\Biggl( \hat{1} + \frac{\alpha_s (\Lambda_1) }{4 \pi} \hat{J} \Biggr)
\hat{U}_{N_f}^{(0)} (\Lambda_1 , \Lambda_2 )
\Biggl( \hat{1} - \frac{\alpha_s (\Lambda_2) }{4 \pi} \hat{J} \Biggr)
,
\end{equation}
where $N_f$ is the number of active quark flavors between the scales $\Lambda_1$ and $\Lambda_2 $.
The matrix $\hat{U}_{N_f}^{(0)} (\Lambda_1 , \Lambda_2 )$ is the evolution matrix of the Wilson coefficients of $|\Delta S| = 1$ four-quark interactions at the leading logarithmic order.
It is given by
\begin{equation}
\hat{U}_{N_f}^{(0)} (\Lambda_1 , \Lambda_2 )
=
\hat{V}
\left(
\begin{array}{cccccc}
\alpha_{12}^{a_1}&&&&&\cr
&\alpha_{12}^{a_2}&&&&\cr
&&\alpha_{12}^{a_3}&&&\cr
&&&\alpha_{12}^{a_4}&&\cr
&&&&\alpha_{12}^{a_5}&\cr
&&&&&\alpha_{12}^{a_6}\cr
\end{array}
\right)
\hat{V}^{-1}
,
\end{equation}
where $\alpha_{12} \equiv \frac{\alpha_s (\Lambda_2)}{\alpha_s (\Lambda_1)} $.
The matrix $\hat{V}$ is the regular transformation matrix which diagonalizes the (transposed) leading logarithmic order anomalous dimension matrix $\hat{\gamma}_{N_f}^{(0)}$:
\begin{equation}
\hat{V}^{-1} \hat{\gamma}_{N_f}^{(0) {\rm T}} \hat{V}
= 
{\rm diag} (E_1, E_2, E_3, E_4, E_5, E_6)
,
\end{equation}
where $E_i$'s are the eigenvalues of $\hat{\gamma}_{N_f}^{(0)}$.
The exponents are $a_i = \frac{E_i}{2\beta_0}$  $(i=1 \sim 6)$.
The leading logarithmic order anomalous dimension matrix of the Wilson coefficients of $|\Delta S| = 1$ four-quark operators is given by \cite{buras,burasreview}
\begin{equation}
\hat{\gamma}_{N_f}^{(0)}
=
\left(
\begin{array}{cccccc}
-2 & 6 & 0 & 0 & 0 & 0 \cr
6 & -2 & -\frac{2}{9} & \frac{2}{3} & -\frac{2}{9} & \frac{2}{3} \cr
0 & 0 & -\frac{22}{9} & \frac{22}{3} & -\frac{4}{9} & \frac{4}{3} \cr
0 & 0 & 6 - \frac{2}{9} N_f & -2+\frac{2}{3} N_f & -\frac{2}{9} N_f & \frac{2}{3} N_f \cr
0 & 0 & 0 & 0 & 2 & -6 \cr
0 & 0 & -\frac{2}{9} N_f & \frac{2}{3} N_f & -\frac{2}{9} N_f & -16 + \frac{2}{3} N_f \cr
\end{array}
\right)
.
\label{eq:leadinganomalousdimension}
\end{equation}

The matrix $\hat{J}$ yields the next-to-leading order correction to the evolution of the Wilson coefficients, and is also defined by using the diagonalizing matrix $\hat{V}$:
\begin{equation}
\hat{J} = \hat{V} \hat{S} \hat{V}^{-1}
,
\end{equation}
where the matrix $\hat{S}$ is defined by
\begin{equation}
\hat{S}_{ij} 
\equiv
\frac{\beta_1}{2 \beta_0^2} \delta{ij} E_i 
- \frac{(\hat{V}^{-1} \hat{\gamma}^{(1){\rm T}} \hat{V} )_{ij}}{2 \beta_0 + E_i -E_j}
,
\end{equation}
with $\hat{\gamma}^{(1)}$ the next-to-leading order anomalous dimension matrix of the evolution of the  Wilson coefficients of the $|\Delta S| =1$ operators.
We must note that $\hat{\gamma}^{(1)}$ is renormalization scheme dependent.
In the naive dimensional regularization scheme, it is given by \cite{buras,burasreview}
\begin{eqnarray}
\hat{\gamma}_{N_f}^{(1)}
=
\left(
\begin{array}{cccccc}
-\frac{21}{2}-\frac{2}{9} N_f & \frac{7}{2}+\frac{2}{3} N_f & \frac{79}{9} & -\frac{7}{3} & -\frac{65}{9} & -\frac{7}{3} \cr
\frac{7}{2}+\frac{2}{3} N_f & -\frac{21}{2}-\frac{2}{9} N_f & -\frac{202}{243} & \frac{1354}{81} & -\frac{1192}{243} & \frac{904}{81} \cr
0 & 0 & -\frac{5911}{486} + \frac{71}{9} N_f & \frac{5983}{162} +\frac{1}{3} N_f& -\frac{2384}{243}-\frac{71}{9} N_f & \frac{1808}{81} -\frac{1}{3} N_f \cr
0 & 0 & \frac{379}{18} + \frac{56}{243} N_f & -\frac{91}{6} +\frac{808}{81} N_f & \frac{130}{9} -\frac{502}{243} N_f & -\frac{14}{3}+\frac{646}{81} N_f \cr
0 & 0 & -\frac{61}{9} N_f & -\frac{11}{3} N_f & \frac{71}{3} +\frac{61}{9} N_f & -99 +\frac{11}{3} N_f \cr
0 & 0 & -\frac{682}{243} N_f & \frac{106}{81} N_f & -\frac{225}{2} +\frac{1676}{243} N_f & -\frac{1343}{6} + \frac{1348}{81} N_f \cr
\end{array}
\right)
.
\nonumber\\
\end{eqnarray}
The scheme dependence of $\hat{\gamma}^{(1)}$ is cancelled by the correction to the operator array
\begin{equation}
\bar{{\mathbf Q}} (\Lambda_1) 
=
\Biggl( \hat{1} -\frac{\alpha_s (\Lambda_1 )}{4 \pi } \hat{r} \Biggr) 
{\mathbf Q} (\Lambda_1)
,
\end{equation}
where $\hat{r}$ is given by \cite{buras,burasreview}
\begin{equation}
\hat{r}
=
\left(
\begin{array}{cccccc}
\ \ \  \frac{7}{3} \ \ \  & \ \  -7 \ \ \  & 0 & 0 & 0 & 0 \cr
-7 & \frac{7}{3} & \frac{2}{27} & -\frac{2}{9} & \frac{2}{27} & -\frac{2}{9} \cr
0 & 0 & \frac{67}{27} & -\frac{67}{9} & \frac{4}{27} & -\frac{4}{9} \cr
0 & 0 & -7 + \frac{5}{27} N_f & \frac{7}{3}-\frac{5}{9} N_f & \frac{5}{27} N_f & -\frac{5}{9} N_f \cr
0 & 0 & 0 & 0 & -\frac{1}{3} & 1 \cr
0 & 0 & \frac{5}{27} N_f & -\frac{5}{9} N_f & -3+\frac{5}{27} N_f & \frac{35}{3} - \frac{5}{9} N_f \cr
\end{array}
\right)
,
\end{equation}
in the naive dimensional regularization scheme.

Finally, the quark threshold matching matrix $\hat{R}$ is given by \cite{buras}
\begin{equation}
\hat{R}
=
\frac{5}{27}
\left(
\begin{array}{cccccc}
0 & 0 & 0 & 0 & 0 & 0 \cr
0 & 0 & 0 & 0 & 0 & 0 \cr
0 & 0 & 0 & 0 & 0 & 0 \cr
0 & 0 & 1 & -3 & 1 & -3 \cr
0 & 0 & 0 & 0 & 0 & 0 \cr
0 & 0 & 1 & -3 & 1 & -3 \cr
\end{array}
\right)
.
\end{equation}

\section{PCAC reductions of pseudoscalar matrix elements \label{sec:pcac}}

The pseudoscalar meson matrix elements can be calculated using the partially conserved axial current formula.
The needed ones are given as
\begin{eqnarray}
\langle \pi^0 | \bar d \gamma_5 d | 0 \rangle
&\approx &
- \frac{i}{f_\pi} \langle 0 | \bar q q | 0 \rangle
,
\label{eq:pi0vacuum}
\\
\langle \pi^+ | \bar u \gamma_5 d | 0 \rangle
&\approx &
\frac{i \sqrt{2} }{f_\pi} \langle 0 | \bar q q  | 0 \rangle
,
\label{eq:pi+vacuum}
\\
\langle K^- | \bar s \gamma_5 u | 0 \rangle
&\approx &
\frac{i}{f_K \sqrt{2} } \langle 0 | \bar q q + \bar s s | 0 \rangle
,
\label{eq:k-vacuum}
\\
\langle \bar K^0 | \bar s \gamma_5 d | 0 \rangle
&\approx &
\frac{i}{f_K \sqrt{2} } \langle 0 | \bar q q + \bar s s | 0 \rangle
,
\label{eq:k0vacuum}
\\
\langle \eta | \bar d \gamma_5 d | 0 \rangle
&\approx &
\frac{i}{f_\eta \sqrt{3}} \langle 0 | \bar q q | 0 \rangle
,
\label{eq:etaddvacuum}
\\
\langle \eta | \bar s \gamma_5 s | 0 \rangle
&\approx &
- \frac{2i }{f_\eta \sqrt{3}} \langle 0 | \bar s s  | 0 \rangle
,
\label{eq:etassvacuum}
\end{eqnarray}
where $q$ represents the light quark.
From the Gell-Mann-Oakes-Renner relation, we have the light quark chiral  condensate $\langle 0 | \bar q q | 0 \rangle = -\frac{m_\pi^2 f_\pi^2}{m_u+m_d}$.
That of the strange quark is given by $\langle 0 | \bar s s | 0 \rangle \approx \langle 0 | \bar q q | 0 \rangle$.

\section{The Chiral lagrangian\label{sec:chiral_lagrangian}}

The chiral $SU(3)_L \times SU(3)_R$ effective lagrangian is given by
\begin{eqnarray}
{\cal L}_0 &=& \frac{f_\pi^2}{4} {\rm Tr} (\partial_\mu U \partial^\mu U^\dagger ) + {\rm Tr} [\bar B (i\partial \hspace{-.6em}/ - m )B] \nonumber\\
&& + \frac{i}{2} {\rm Tr} [\bar B \gamma_\mu (\xi \partial^\mu \xi^\dagger +\xi^\dagger \partial^\mu \xi)B]
+ \frac{i}{2} {\rm Tr} [\bar B \gamma_\mu B (\partial^\mu \xi \xi^\dagger +\partial^\mu \xi^\dagger \xi)] \nonumber\\
&&+ \frac{i}{2} (D+F) {\rm Tr} [\bar B \gamma_\mu \gamma_5 (\xi \partial^\mu \xi^\dagger -\xi^\dagger \partial^\mu \xi)B ]
\nonumber\\
&&- \frac{i}{2} (D-F) {\rm Tr} [\bar B \gamma_\mu \gamma_5 B (\partial^\mu \xi \xi^\dagger -\partial^\mu \xi^\dagger \xi) ] \ ,
\label{eq:chiral_lagrangian}
\end{eqnarray}
with $B$ the baryon field.
The meson field $M$ is introduced through
\begin{equation}
\xi = \exp \left( \frac{iM}{\sqrt{2}f_\pi } \right) \ ,
\end{equation}
with $U=\xi^2$.
The meson matrix is given by
\begin{equation}
M=
\left(
\begin{array}{ccc}
\frac{\pi^0}{\sqrt{2}}+\frac{\eta}{\sqrt{6}}&\pi^+&K^+\\
\pi^- &-\frac{\pi^0}{\sqrt{2}}+\frac{\eta}{\sqrt{6}}&K^0 \\
K^- & \bar K^0 & -2 \frac{\eta}{\sqrt{6}}
\end{array}
\right) \ ,
\end{equation}
and the baryon matrix by
\begin{equation}
B=
\left(
\begin{array}{ccc}
\frac{\Sigma^0}{\sqrt{2}}+\frac{\Lambda^0}{\sqrt{6}}&\Sigma^+&p\\
\Sigma^- &-\frac{\Sigma^0}{\sqrt{2}}+\frac{\Lambda^0}{\sqrt{6}}&n \\
\Xi^- & \Xi^0 & -2 \frac{\Lambda^0}{\sqrt{6}}
\end{array}
\right) \ .
\end{equation}

The zeroth order expansion of the third line of Eq. (\ref{eq:chiral_lagrangian}) in the meson field can be transformed as 
\begin{eqnarray}
{\cal L}_{(M^1)} &=& \frac{D+F}{\sqrt{2} f_\pi} {\rm Tr}\, [ \bar B \gamma_\mu \gamma_5 (\partial^\mu M ) B]
+\frac{D-F}{\sqrt{2} f_\pi} {\rm Tr}\, [ \bar B \gamma_\mu \gamma_5 B (\partial^\mu M )] 
.
\label{eq:chiral_lagrangian2}
\end{eqnarray}
Terms without nucleon field operators were omitted since they do not contribute to the CP-odd nuclear force in the one-meson exchange model.
By considering the on-shell contribution of the baryon as leading, Eq. (\ref{eq:chiral_lagrangian}) can be approximated by using the equation of motion $\bar B \gamma_\mu \gamma_5 B' \partial^\mu M \approx -i (m_B + m_{B'})\bar B \gamma_5 B' M$.
We therefore obtain the following lagrangian:
\begin{eqnarray}
{\cal L}_{mBB} 
&\approx & 
-g_{\pi NN} 
\bigl[
\bar pi\gamma_5 p \pi^0 
-\bar n i\gamma_5 n \pi^0 
\bigr]
- \sqrt{2}g_{\pi NN} \bigl[ 
\bar p i\gamma_5 n \pi^+ +{\rm h.c.}
\bigr]
\nonumber\\
&&
-g_{\eta NN} 
\bigl[
\bar p i\gamma_5 p \eta 
+\bar n i\gamma_5 n \eta 
\bigr]
\nonumber\\
&&
+g_{K\Lambda N} 
\Bigl[ 
\bar p i\gamma_5 \Lambda K^+ 
+\bar n i\gamma_5 \Lambda K^0 
+ {\rm h.c.}
\Bigr]
\nonumber\\
&&
+g_{K\Sigma N} 
\Biggl[
\bar \Sigma^+ i\gamma_5 p \bar K^0 
+ \bar \Sigma^- i\gamma_5 n K^-
+\frac{1}{\sqrt{2}} 
\Bigl(
\bar \Sigma^0 i\gamma_5 p K^-  
-\bar \Sigma^0 i\gamma_5 n \bar K^0
\Bigr)
+ {\rm h.c.}
\Biggr]
,
\ \ \ \ \ \ 
\label{eq:chiral_meson-baryon}
\end{eqnarray}
where
$g_{K\Lambda N}=\frac{m_N+m_\Lambda}{2\sqrt{3} f_\pi } (D+3F) \approx 13.6$ and
$g_{K\Sigma N } = -\frac{m_N +m_\Sigma}{\sqrt{2} f_\pi } (D-F) \approx - 6.0$, with $D=0.81$ and $F=0.44$.
Here we have assumed the isospin symmetry.
Note that the sign is different for terms involving kaons.
In this chiral approach, the pion-nucleon and eta-nucleon couplings are given by $g_{\pi NN } = \frac{m_N }{f_\pi } (D+F) \approx 12.6$ and $g_{\eta NN } =\frac{m_N }{\sqrt{3}f_\pi } (3F-D)\approx 3.0$, respectively, but we use phenomenological data \cite{ericson,tiator} as input in this work.


\end{document}